# A Co-design view of Compute in-Memory with Non-Volatile Elements for Neural Networks

*Wilfried Haensch[1,*], Anand Raghunathan[2], Kaushik Roy[2], Bhaswar Chakrabarti[3], Charudatta M. Phatak[1], Cheng Wang[2] and Supratik Guha[1,4,*]*

(1) Materials Science Division, Argonne National Laboratory

(2) Purdue University, Department of Electrical Engineering, West Lafayette, IN 47907

(3) Department of Electrical Engineering, Indian Institute of Technology Madras, Chennai, Tamil Nadu 600036, India

(4) Pritzker School of Molecular Engineering, University of Chicago, Chicago, IL 60637

**\*** Corresponding authors: whaensch@anl.gov, sguha@anl.gov

**Abstract:**
Deep Learning neural networks are pervasive, but traditional computer architectures are reaching the limits of being able to efficiently execute them for today's large workloads. They are limited by the von Neumann bottleneck: the high cost in energy and latency incurred in moving data between memory and the compute engine. Today, special CMOS designs address this bottleneck. The next generation of computing hardware will need to eliminate or dramatically mitigate this bottleneck. We discuss how compute-in-memory can play an important part in this development. Here, a non-volatile memory based cross-bar architecture forms the heart of an engine that uses an analog process to parallelize the matrix vector multiplication operation, repeatedly used in all neural network workloads. The cross-bar architecture, at times referred to as a neuromorphic approach, can be a key hardware element in future computing machines. In the first part of this review we take a co-design view of the design constraints and the demands it places on the new materials and memory devices that anchor the cross-bar architecture. In the second part, we review what is knows about the different new non-volatile memory materials and devices suited for compute in-memory, and discuss the outlook and challenges.

# Contents

Table of Acronyms

| | | | |
|---|---|---|---|
| ADC | Analog-to-Digital Converter | MANN | Memory-Augmented Neural Network |
| AI | Artificial Intelligence | ML | Machine Learning |
| ANN | Artificial Neural Network | MLP | Multi-Layer Perceptron |
| ALU | Arithmetic and Logic Unit | MRAM | Magnetic Random-Access Memory |
| ASIC | Application Specific Integrated Circuit | MTJ | Magnetic Tunnel Junction |
| CIM | Computing In Memory | MVM | Matrix-Vector Multiplication |
| CMOS | Complementary Metal-Oxide-Semiconductor | NVM | Non-Volatile Memory |
| CNN | Convolutional Neural Network | PCB | Printed Circuit Board |
| CPU | Central Processing Unit | PCM | Phase Change Memory |
| DAC | Digital-to-Analog Converter | PE | Process Engine |
| DL | Deep Learning | RRAM | Resistive Random-Access Memory |
| DNN | Deep Neural Network | SGD | Stochastic Gradient Descent |
| DRAM | Dynamic Random-Access Memory | SNN | Spiking Neural Network |
| GNN | Graph Neural Network | SOT | Spin Orbit Torque |
| GPU | Graphic Processing Unit | SRAM | Static Random-Access Memory |
| LIF | Leaky Integrate and Fire | STT | Spin Transfer Torque |
| LSTM | Long-Short Term Memory network | STDP | Synaptic Time Dependent Plasticity |
| MAC | Multiply-and-Accumulate operation | TPU | Tensor Processing Unit |
| | | $V_{mem}$ | Membrane potential |

# 1 Introduction

The field of Artificial Intelligence (AI) and in particular, Machine learning (ML), has witnessed tremendous progress in recent decades. This is in large part due to the (re-)emergence of neural networks, fueled by the availability of efficient and affordable computing hardware to build models utilizing back propagation, as the algorithm of choice for an ever-increasing range of machine learning tasks. From its very early days, the field of AI has aimed to emulate aspects of natural or human intelligence. This influence can be traced back to the work of George Boole [1], John von Neumann [2] and Alan Turing [3] that laid the underpinnings of digital computing. Within the field of AI, notable examples of brain-inspired advances include artificial neural networks (ANNs), which are comprised of artificial neurons and synapses; deep learning neural networks, which are ANNs with several layers that lend to automatically learned feature representations; reward-driven learning, which underpins reinforcement learning and attention; episodic and working memory, and many more [4].

The past decade in particular has witnessed meteoric progress in the field of AI due to a virtuous cycle between algorithms, computing platform capabilities, and proliferation of data. The availability of large, **open access** datasets, particularly for computer vision [5] [6], natural language [7] and speech processing [8] and the emergence of highly parallel computing platforms (both at the chip scale, such as GPUs, and at the data center scale) provided machine learning researchers with an ability to create models of unprecedented complexity and accuracy, at times surpassing human capabilities for specific applications. In turn, the success of deep neural networks in many applications and the availability of larger amounts of data has driven a seemingly insatiable demand for computational resources. As an example, the computation required to train the largest models is estimated to double every 3.5 months [9].

As transistor technology has scaled, improvements in ML hardware performance have been increasingly delivered by innovations in heterogeneous integration and specialized ML accelerators (Google's TPU, NVIDIA's Tensor Cores, IBM Tellum, Intel Habana, Graphcore, Cerebras are examples). Most of these digital accelerators feature specialized data-paths (regular arrays of multiply-accumulate processing elements for matrix vector multiplications); and memory hierarchies that are well-matched to the patterns observed in deep neural networks, and employ approximate (e.g., lower precision) hardware to perform computations. The trends in ML workloads that will drive the next generation of hardware are both qualitative and quantitative. As deep neural networks proliferate into a wider range of applications, and as researchers pursue improved capabilities within each application domain, the network topologies have not only grown more complex (deeper networks, more parameters) but also more varied. Therefore, future computing platforms may need to execute different classes of neural networks, including multi-layer perceptrons (MLPs) [10], convolutional neural networks (CNNs) [11], long-short term memory networks (LSTMs) [12], Transformers, memory-augmented neural networks (MANNs) [13], graph neural networks (GNNs) [14], and possibly more. Finally, there are efforts towards the next wave of brain-inspired concepts, such as spiking neural networks (SNN), localized learning algorithms, and high-dimensional distributed information representations. These trends in algorithms and workloads can profoundly alter the demands on hardware. In addition, the increase in model and data sizes will continue to drive the demand for increased storage and computation.

It is highly unlikely, if not inconceivable, that simply continuing the current trends (larger and deeper networks, reduced precision) can keep up with the scale and flavor of future machine learning workloads. Almost all of the workloads and classes of neural networks noted above need to execute linear algebraic operations that involve multiple matrix vector multiplications (MVM) that involve large weight matrices. The total number of weights, or learned parameters, can exceed hundreds of millions; for instance OpenAI LLC's GPT-3 natural language processing model, has 175 billion parameters [15].  As will be described later, carrying out these multiplications using current digital CMOS hardware is energetically very expensive because of repeated data movement between memory and the logic processors. This energy cost is daunting in light of the anticipated trends in workload growth.  Emerging hardware fabrics based on a cross-bar architecture and non-volatile memory elements lead to compute-in-memory approaches that can potentially provide a large boost in processing speed and energy efficiency by carrying out these matrix vector multiplications in analog mode (described later). **These cross-bar architectures are yet at a research stage, but should they be technologically viable, they have the potential to be one of the keystone hardware elements within future processor architectures for ML workloads**. **We focus on this cross-bar based compute-in-memory approach, which embeds computational capabilities into memory arrays to overcome the data movement bottleneck in current hardware platforms**. In this review, we take a co-design view: we explore the role of new materials and devices in addressing this challenge, but in the context of the demands that micro-architectural, circuit and technology relevancies place upon these materials.

Deep learning (DL), driven by DNNs has gained popularity since its re-emergence in 2012 [5]. A DNN network consists of a number of layers containing the "synaptic" weights [16]. These layers can be visualized as two-dimensional arrays, represented by matrices, in which the weights are encoded in the matrix elements or the cross-points. Layers are connected through a non-linear function (activation function) to covert the output from the previous layer to the input for the next layer (activation). There are two operation modes: (1) Inference and (2) Training. In the latter, a model is built where the weights are adjusted and optimized based on existing data and known results. In the former, this model (with fixed weights) is then used to predict unknown results (such as classifying unknown objects) using available data. The details of the training process (see ref. [16] for details on this) are not important for what follows. It consists of three parts [10]:

>(1) Forward-path: The input data is prepared as a vector (for instance the columns of a 2D pixel array are stacked to form a 1D vector) and fed into the network where it moves from layer to layer, in a chain of matrix-vector multiplications (MVMs), until it reaches the end of the network. For training the input vector x for each layer is stored and the output of the last layer is compared with the expectation (based on the known input data) to calculate an error vector $\delta$. In the case of inference, no intermediate result is kept and the output of the last layer is the classification of the input data.

>(2) Backward-path: This operation mode only occurs for training. The calculated error $\delta$ moves backward through the network in a chain of MVMs from the end to the beginning and the intermediate results $\delta$ are stored.

>(3) Update: During this phase the stored data from the forward-path and the backward-path is recalled for each individual layer to update the weights.

This process is repeated many times until the desired error rate is reached. During the training phase the complete dataset (epoch) is presented to the network many times which makes training a highly compute intensive process, especially considering that modern datasets can have many millions of entries. In inference the unknown data is presented to the trained model and the signal travels forward through the net and results in a categorization at the end. This process is identical with the step (1) in the training process. This makes inference significantly inexpensive computationally compared to training. Mathematically steps (1) and (2) are represented by MVMs. An MVM will have $N^2$ multiply and accumulate (MAC) operations, or $2N^2$ floating point operations (Op), if the input and output vectors are of dimension $N$. In step (3) the weight update is accomplished by an incremental change of the weight matrix element $w_{i,j} \leftarrow w_{i,j} - \eta x_i \delta_j$ with $x$ the input vector from the forward (activation) pass and $\delta$ the error vector from the backward pass and $\eta$, a scale factor to avoid large changes. This implies, $N^2$ MAC operations as well (for an $N \times N$ matrix) in step 3. The goal of computer architectures for DL is to make these operations as energy efficient as possible. While the forward (1) and backward (2) process is a simple MVM on stationary weights, the update operation (3) manipulates the weights itself. However, all three operations have as a basic ingredient the MAC operation. From a computational complexity point of view (1) & (2) require only stationary weights. For the update operation (3), however, the system has to keep track of the weights, activations and errors at any time during the training process and therefore training is considerably more taxing on the memory requirements than inference. Therefore, training is typically relegated to data centers while inference finds its way into consumer products or towards the edge in applications such as IoT (internet-of-things). Technology objectives for both application spaces are therefore different. While inference applications strife for power efficiency measured in operations per Joule (Op/J) the training objective is mostly to reduce the training time or frames per second (F/s).

To increase compute efficiency for ML workloads there are three principal approaches that are often referred to a "compute-in-memory" or "compute-near-memory":

a) ***Compute-near-memory: Digital memory and digital compute (Training & Inference)***: This is at present the most popular approach in the industry and is entirely carried out using conventional digital CMOS approaches. Bringing memory closer to the compute engine and deploying reduced precision representation of the digital content are common goals for architecture design here (discussed further in Section 2).
b) ***Compute-in-memory: Digital memory and analog compute (Inference):*** This approach can significantly increase parallelism and reduce data transport to and from off-chip memory. Memory elements storing the weights, in one bit or multibit capacity, are arranged at the nodes of a crossbar array of metal interconnects. Here, the MVM operation is carried out in analog fashion using Kirchhoff's and Ohm's laws: a voltage vector is applied at the rows of the array and the current is integrated at each column. A fixed integration time, $t_{int}$, is a design parameter of the hardware configuration. Which means that $2N^2$ operations are performed within this integration time, if the matrix is of the size *N x N*.
c) ***Compute-in-memory: Analog memory and analog compute (Training & Inference):*** Here, the weights at the cross-bar nodes are encoded in memory elements that take on analog values along a sliding continuous scale between a high and low conductance value. The MAC operation is similar to that of (b). This mode of operation also enables incremental changes of all the array elements in parallel at constant time, usually of the order of the integration time. This approach (as well as (b)) requires analog to digital conversion peripheral electronics to communicate with the digital computing environment for further processing of the array output.

Approach (a) is a relatively mature area with significant commercial success. Innovations are carried forward via CMOS design and heterogeneous integration approaches in order to increase speed and energy efficiency. Approaches (b) and (c) can provide further gains in efficiency due to parallelization and reduction in the memory bottleneck, but require considerable improvements in materials and device performance in order to have viable commercial impact. The topics for our subsequent discussions will be largely restricted to items (b) and (c). In the discussion below it will become clear that material requirements for options (b) and (c) will be different in significant details, see Section 2 and Section 3 respectively. Applications (b) and (c) are often also referred to as "neuromorphic computing". The cross-point memory elements are sometimes referred to as synapses and the peripheral components to the array that provide the "activation function", as neurons.

Beyond the hardware approaches for DL described above, spiking neural networks (SNN) have drawn considerable interest because of their perceived proximity to realistic biological systems. In SNNs the communication between the components is in the form of temporally sparse voltage pulses (spikes) instead of a well-defined digital signal. These spikes are launched when a neuron (activation) exceeds a threshold and fires a signal that propagates to the next layer. As a result, communication channels are less frequently used than in a continuous digital flow which results in power savings. In contrast to DL systems, in biological systems (and SNNs) the signal can travel in both direction at the same time in a specific layer which introduces a temporal component to the signaling. The update of the weights now depends on the relative sequence of these signals and is described by synaptic spike timing dependent plasticity (STDP) [17]. This is very different that the static weight update in the back propagation algorithm. Training of these systems is increasingly difficult since they cannot rely on an error that is passed from a target function traveling back through the system. While SNNs may offer advantages [18], their successful deployment is a considerably more complex task than the current compute-in-memory approaches to DL.

The paper is organized as follows: After discussing the motivation for compute in-memory in Section 2 we will focus on inference applications for compute in-memory (CIM) with arrays of NVM representing the weight matrices of the DL network. Here we will discuss how arrays of memory elements can be used to accelerate the MVM that is frequently used in DL algorithms. In accomplishing analog MVM we will distinguish the use of the memory elements as digital storage nodes with one or multiple bit capacity and the analog counter part in which the information is stored on a continuos scale. In both, the memory element at each each cross-point performs a multiply-accumulate (MAC) operation simultaniously resulting in a high degree of parallel operations. We describe solutions for their implemenation and the challenges to overcome. Finaly we extend the discussion to the all analog scenario where the weight update is also performed in parallel in these arrays. The core of this paper is Section 3 in which we discuss the choice of materials and their level of maturity for use in these accelerator arrays. We discuss the relevance of the materials for memory application as well as their potential to be used for compute-in-memory (CIM). We describe switching mechanisms, material choices, level of maturity and challenges that might hinder implementation. A discussion of our findings is presented in Section 4. A short conlusion in Section 5 and a summary table of the current state of the art materials is at the end of this review.

## 2 Compute-in Memory, hardware considerations
### 2.1 Motivation

MAC units used for MVMs are the major processing elements for most machine learning algorithms. It is also important for HPC systems which increasingly exploit GPUs for large-scale vectorized matrix arithmetic operations. When ML algorithms are implemented in a standard von Neumann architecture, such as CPUs or GPUs, the data movement cost (in energy and latency) for fetching weights and activations from the memory (on-chip cache and off-chip main memory), transferring them to the processing units, and storing the computations back to the memory can be significant. Over the years, memory latency (time between requesting the data and its arrival time) has not kept up with processor performance, in spite of a steady progress in memory bandwidth (how much data can be transferred in a second) [19]. This creates a performance gap for data rich applications. The bottleneck arising from such data movements between memory and the processing unit is referred to as the "von Neumann memory wall", as is shown in Figure 1. Bringing memory on the same chip as the compute unit in the form of cache memory addresses this problem in part but cannot avoid it.

Cache memories are usually fast SRAM (at a lower density) compared to the slower, but higher density DRAM which is used as off-chip memory. It is worthwhile to note that accessing off-chip DRAM requires approximately two orders of magnitude more energy than the on-chip SRAM access [20], [21] and about an order-of-magnitude more in latency.

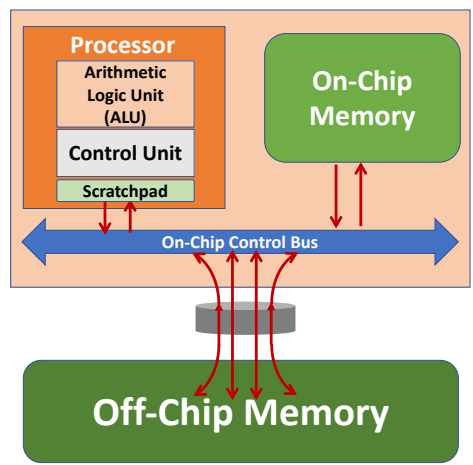

**Figure 1:** Schematic of a von Neuman computer architecture. A controller regulates data and instruction flow from memory to the arithmethic and logic unit (ALU) for processing. On chip memory (or cache) helps to reduce off-chip traffic.

To address the high cost of data movement, general architectures (such as GPUs) bring distributed on-chip memory closer to processing units, particularly to frequently used processing units such as multiply-and accumulate (MAC) engines. As is shown in Figure 2 (left), such compute "near-memory" allocates some small on-chip cache memory to each processor unit in order to mitigate the memory bottleneck. However, each processor in a GPU *still* needs to fetch data from the dedicated memory to the compute unit, execute the compute operation (such as multiply and accumulate) and then write back to the local memory. Therefore, while the proximity of the memory units to the processors can alleviate the "memory wall" bottleneck, it still remains (due to the performance gap between the processor and the memory). This is a limitation of current CMOS based compute near-memory. The limitation is compounded by the main types of memory available today: DRAM and SRAM. Efforts aim at reducing the off-chip memory access by increasing the on-chip memory capacity. Dynamic random-access-memories (DRAMs) are the main off-chip memory: While they offer high memory densities (with a cell structure of 1-transistor-1-capacitor, 1T1C) and are cheaper, they are limited in access speed (50-100ns), and need to be built on a separate chip—since DRAM technology is difficult to integrate with logic technology. Consequently they are not appropriate for on-chip memory and and are connected to the processor chip via lengthy interconnects. These will be discussed briefly later, but the overall impact is that DRAM is energetically very expensive for MAC operations. DRAM is typically designed as the main (off-chip) memory in modern computing hardware. In order to cope with the high processing speed of computing units, on-chip memories need to have high density, high bandwidth, high reliably, and offer fast and energy-efficient read/write operations to minimize the off-chip memory access cost. The six-transistor static random access memory (SRAM) meets these requirements and is the standard on-chip memory used today. While read and write operations of SRAM can be fast (< 1ns access time) and efficient (<0.1pJ write energy), the density of SRAM (which includes six transistors in its design) is significantly limited due to its large area footprint (> $140F^2$, F being the feature size). The resulting low density of SRAM is a significant drawback for its use as an on-chip memory for MAC engines and other applications. More recently embedded DRAM (eDRAM) has been used in higher level on-chip cache memory [22] . Reaching an SRAM cache level speed with eDRAM, however, is at the cost of a larger cell size that can be as large as $50F^2$ which still provides a density advantage to SRAM. Both SRAM and DRAM are volatile, meaning the stored bits will be lost once the power is off. The volatility of memory also contributes significantly to the leakage power at advanced nodes. In view of the aforementioned "memory bottleneck", the limitations of DRAM and SRAM, and the need to reduce the data movement between the main memory and on-chip caches, compute-in-memory approaches have been proposed (Figure 2, right) that use a cross-bar architecture and NVM memory elements to accomplish the MAC operation in analog rather than digital fashion. This is discussed in the next section.

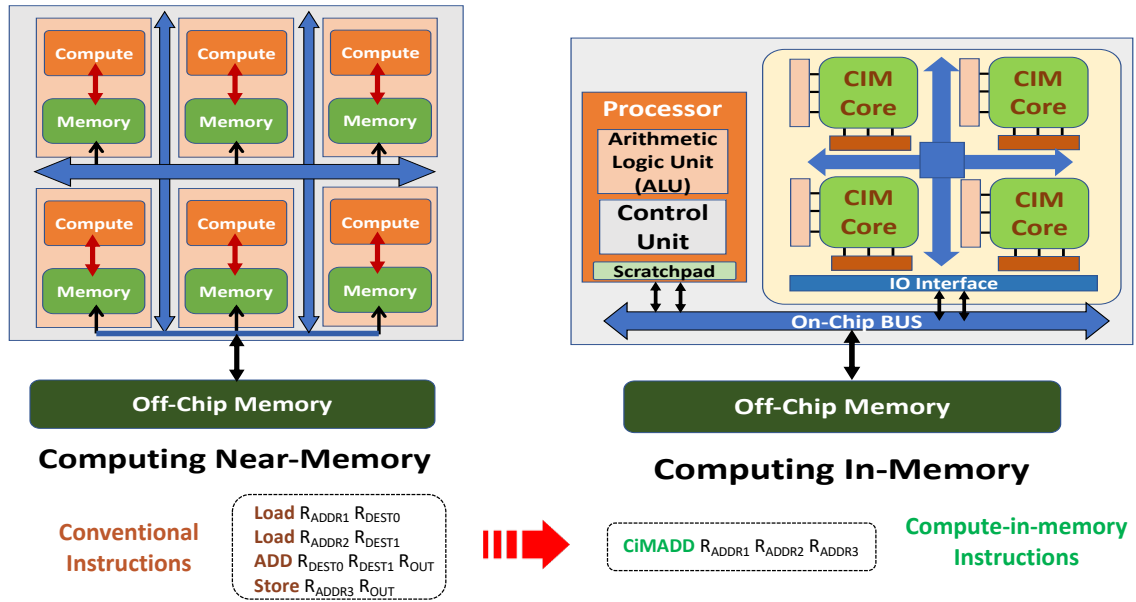

**Figure 2:** Left - compute near memory: Compute and memory are in close proximity but still separate units. Load instructions are used to fetch data from the memory to the compute unit, in which the computation (for example an addition) is performed and the results are stored back to memory. Registers ($R_i$) are used to specify the memory addresses as well as the instruction operands. Right – compute in memory: Compute and memory are indistinguishable units, reducing the data flow. In this case, a single instruction (CiMADD) is used in which the computation as well as the addresses of the operands and result are sent to memory, saving data transfers between the memory and compute. Note, in both cases there is still a need to access external off-chip memory.

## 2.2  The cross-bar architecture as an analog MAC engine

In order to further address the memory bottleneck, computing in-memory (CIM) has been proposed by merging on-chip compatible memory and compute units into CIM cores formed by crossbar arrays [23], further reducing the data movements between off-chip-memory, on-chip cache and compute unit for MVM operations.  Figure 3 shows a cross-bar structure and its associated hardware layout. As an example, a chip (Figure 3a) is composed of a number tiles that contain the cross-bar arrays to perform MVM operations; and a digital compute unit for other operations. The tiles and the digital compute unit communicate (digitally) via a data bus. I/O circuitry provide the means to communicate off-chip. Other tile architectures and communication between them are discussed as well [24] [25]. Each tile contains the cross-bar memory array along with ADC, and control/communications digital circuitry (Figure 3b). The crossbar memory array (Figure 3c) stores the weights that are written into the memory elements at the cross-bar nodes (Figure 3d), while access transistors or other selector devices may be present (application dependent) to prevent unintended leakages during write operations [26] [27]. Various memory technologies including SRAM [28] [29] [30] [31] [32]and NVMs [33] [34] [35] [36] [37] have been developed for crossbar hardware, which will be discussed in further detail later. Furthermore, various types of devices including two-terminal selectors [38] and three-terminal transistors have been investigated for the access devices. Access transistors have more symmetric I-V characteristics and less severe device non-linearity, while the 2-terminal selector devices could potentially deliver higher density.

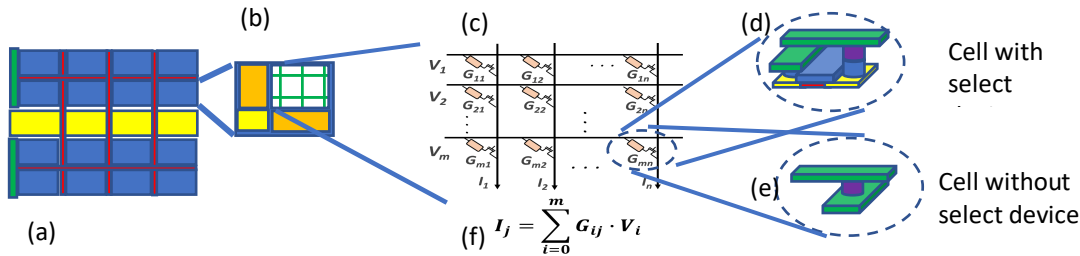

**Figure 3:** A hardware layout based around a cross-bar architecture for matrix-vector-multiplication. (a) Analog components are arranged in a tiled configuration. The tiles (blue) communicate with each other and with other digital circuitry via a bus (red) and with the external world (off-chip I/O, green). Control and additional computation components (yellow) complete the system. (b) The tile containing the cross-bar array (cross-hatch): ADC, activation or array analog activation, digital control for communication and tile I/O; (c) details of the the crosss-bar; (d) the memory cell at the cross-bar nodes that hold the synaptic weights in the memory element (purple), with aselect transistor (d) or without select device (e). During a cross-bar based matrix vector multiplication the input voltage vector $[V_i]$, (i=1,m), applied to the m rows of the array, drives current vector $[I_j]$, (j=1,m), that is collected at the end of column $j$. Ohm's and Kirchhoff's laws are used to achieve equivalency with a vector matrix multiplication operation. As shown in (f), this leads to a multiplication of vectors $[V_i]$ with matrix $[G_{ij}]$, (i=1,m) where $G_{ij}$ are the cross point conductances (also called the synaptic weights, or weights) at the nodes of the array.

Using the cross-bar architecture, matrix vector multiplications (MVM), which dominate most DNN workloads, can now be executed *in-situ* with massive parallelism following Ohm's law (refer to Figure 3c and the description of the operation in the figure caption). The MVM outputs are obtained by integrating the accumulated charge per column (bit lines), converting it into a digital signal (ADC) and then feed the signal to the digital compute unit for further processing. Conceptually, crossbar arrays emulate the synaptic connections between pre-neurons (array input) and post-neurons (array output) in a neural network, where input stimuli from previous layers can propagate to later layers through the interconnected synaptic weights. The CIM cores also contain peripheral circuitry such as input drivers and analog-to-digital converters (ADC) to interface with the digital data, since all communication with other computational components are still in the digital domain. The memory elements at the cross-bar nodes can store weights either digitally, or along a continuous sliding scale (leading to analog storage). Both situations will be considered below.

## 2.3 Considerations for compute-in-memory with digital storage: bit slicing

It is worth noting that the numerical precision of the neural network's weights may be different from the number of bits that can be stored in a single memory device. This may require a weight to be spread out across multiple memory cells [39] [40]. Moreover, even if a single memory cell can store a multi-bit weight, it may lead to an excessively large overhead in the ADCs. Therefore, bit-slicing is commonly used to split a multi-bit weight such that it is stored in multiple crossbar slices. Similarly, bit-streaming is used to process a higher-precision activation by temporally streaming the constituent bits through the crossbars that perform MVM. Specifically, input vectors are converted to bit streaming into low-precision input drivers, while high precision weights are partitioned into bit slices and mapped into multiple low-precision memory devices, as illustrated in the Figure 4.

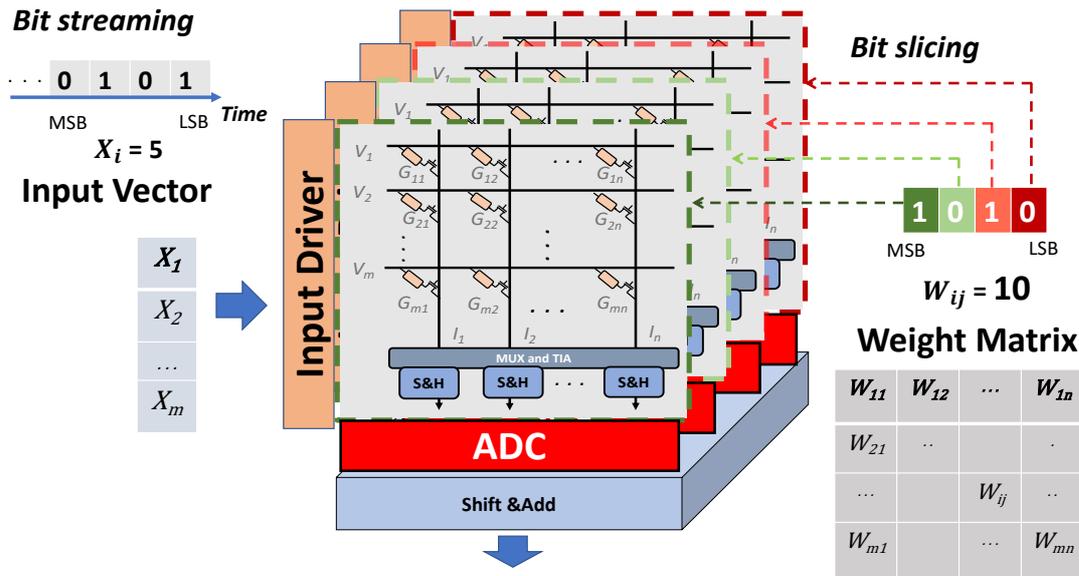

**Figure 4:** Hardware implementation of high-precision input vector and weight matrix using bit streaming and bit slicing. Bit-slicing allows a higher precision weight to be spread out across multiple memory cells. In this example each of the weights in the *M x N* weight matrix, $W_{ij}$, contains 4 bits, while each memory device in the cross-bar array can only store a single bit. Thus, the matrix is stored across 4 crossbar arrays from the most significant bit (MSB, green) in each $W_{ij}$ stored in the first cross-bar, in to the least significant bit (LSB, red) stored in the fourth cross-bar. Bit-streaming is used to present the 4-bit input across 4 time-steps so that only 1 bit is processed at a time. Bit streaming reduces the overhead of digital-analog conversion at the input driver. Each crossbar array performs an analog multiplication receiving the same input vector. The output currents from each crossbar will be converted to digital bits and will be shifted and added to obtain the final MVM**.**

At the end of the crossbar, the accumulated analog currents flowing in the vertical bit lines are first converted to voltages by trans-impedance amplifiers (TIA), and then converted to digital values by the ADC. Sampling-and-hold (S&H) circuits are needed to cope with bit streaming, and shift-and-adders are connected to obtain the high-precision results.

Sensing and differentiating the crossbar currents requires a bit precision (the number of bits produced at the ADC output) of at least ($log_2 M + I + W - 2$) bits to represent the exact results of the MVM operations, where *M* is the crossbar array size, *I* is the number of bits in each input element and W ist he number of bits in each weight. For example, a moderate array size containing 128 rows with 2-bit memory devices and 1-bit per stream already requires a 8-bit ADC [41]. The need for such high-precision ADCs dominates the energy and area of the crossbar hardware [42]. As is illustrated in Figure 5, ADC consumes about 60% of the total energy, and over 80% of the chip area in CIM hardware [27]. In addition, due to the large area, ADC circuitry are typically shared across multiple columns accessed with multiplexers (MUX). Such column-sharing of ADC contributes to a significant latency in MVM operations. Considering the significantly increased energy and area overhead of peripheral input driver/DAC and ADC circuitry at high bit precision, CIM architecture typically uses 1-bit streaming for input encoding, while bit slicing is found to optimize at no more than 4 bits per slice.

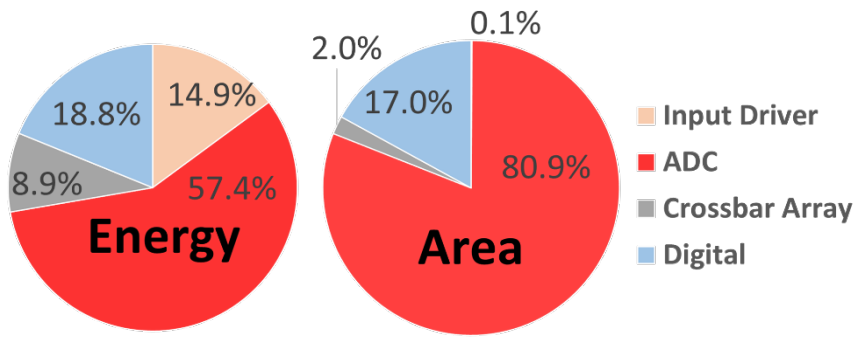

**Figure 5:** ADC dominance in hardware cost. Simulation results based on in-memory crossbar hardware with resistive non-volatile memory. The NVM Crossbars in the hardware has an array size of 128x128 with 2 bits per cell and 1bit per stream. Therefore, 8-bit ADC is required in order to sense the output current of MVM operations

## 2.4   Partitioning of large matrices

In the preceding discussion, we showed how a weight matrix comprising of higher-precision weights could be sliced into multiple crossbar arrays of identical dimensions but lower precision memory cells. However, in general the dimensions of the weight matrix may be larger than the cross-bar array sizes that can be realized. For example, state-of-the-art DNN models contain numerous layers with hundreds of millions to billions of parameters. In contrast, crossbar arrays are typically designed to have $10^3$ to $10^4$ synaptic elements, due to practical device and circuit level constraints [43]. Therefore, large matrices, such as some fully connected layers in the DNN workload will need to be partitioned into multiple arrays as illustrated in Figure 3a.

Furthermore, convolutional operations, which are building blocks of the ubiquitous convolutional neural networks (CNN) in many vision-related tasks, can be converted to iterative MVM operations based on crossbar arrays. It has been demonstrated that large-scale CNN models with large number of kernel channels can be mapped into a crossbar architecture through the tiling of multiple arrays [39] [44]. Following such mapping of convolutional operations, filter weight reuse is achieved and partial MVM results at the array level will be accumulated to generate the final convolutional output.

## 2.5 Challenges and Requirement for designing synaptic devices for crossbars

Due to the analog nature of the MVM operations with crossbar memory arrays, the illustrated *in-situ* computation of MVM is prone to errors caused by various sources of device and circuit non-idealities [43], [45]. To that effect, designing a scalable and reliable on-chip storage of the weight matrices is crucial [46]. In order to understand the device requirement for crossbar-based in-memory compute, we will first discuss the major attributes of these non-idealities.

*IR drop:* Cross-bar hardware, which is based on the sensing of analog currents to obtain the MVM results, is susceptible to parasitic voltage drops ("IR drop") on circuit components including wire resistances, source resistance at the input driver, and source resistance at the connection to sensing circuit. It has been demonstrated that the IR drop leads to significant error due to the reduction of the output currents in the array, especially for larger array sizes [43]. Higher device ON-resistance (LRS) can help mitigate the IR drop, very high ON-resistances give rise to other complications such as sensing challenges. Having a higher ON/OFF ratio of a memory device may also help to mitigate the IR drop impacts.

*Noise and drift*: Besides the conductance range the memory element is working in and its endurance the other material properties that are concerning are: temporal and spatial variations, noise and systematic conductance drift. Spatial variations describe the situation that if all conductances in an array are written with the same stimulus their conductance value $g_{ij}$ will fluctuate around the mean $\tilde{g}$ with a variation $\delta g(g_{ij})$. In the same spirit, if the same element is visited time after time a temporal variation of $g_{ij}$ with variation $\delta g(g_{ij},t)$ is observed. The combination of both will result in a spread of $g_{ij}$ across the array varying with position and time. The same can be said for the incremental change in conductance $\Delta g$, which will directly determine the number of programing steps. The random nature of these variations only allows to specify their mean and tolerances. The individual values will be scattered across the array. The question at hand is: what are the proper means and how large can these fluctuations be without impacting the classification accuracy of the results and how can they be characterized at the individual device level? In addition, a systematic drift $g_{ij}(t)$ on top of these variations will alter the initial weight distribution as well and can degrade the accuracy. Unfortunately, there is not easy way to capture the fluctuation bounds independently of the application. Only their impact on the classification accuracy will provide the tolerable bounds. Therefore, it is important to develop simulation tools that can capture these variations for relevant examples [47].

*Sensing challenges:* The operations of deciphering high-precision MVM output from the sensing of analog crossbar currents places remarkable challenges for the device technology. To that effect, the ON/OFF ratio of each synaptic device, together with the number of bits per cell, in the memory element determines the sensing margin of the output in the crossbar—the voltage window that needs to be sensed accurately to distinguish two adjacent memory states. Moreover, the variation or random fluctuation of device resistance states may lead to additional noise. Note that the impact of device variability on the errors will also become more severe as the number of bits per cell increases. In addition, noises from peripheral circuit such as ADC quantization noises may also contribute to errors during sensing.

*Write challenges:* While ML inference can be implemented with stationary weight mapping which does not require frequent writing, other applications such as ML training and on-chip learning may require weight updates. In such scenarios, the write endurance, programmability of the memory devices, write energy and latency are important parameters for design consideration. In addition, the linearity of conductance modulation under programming pulses is also critical for weight updates. Strong non-linearity in the conductance modulation will require complex lengthy write-verify cycles during weight updates, leading to significant increase of write latency and energy.

The need to minimize the analog computing error due to the non-idealities noted impose certain **design requirements for cross-bars** in order to achieve both hardware and functional accuracy, while retaining high on-chip density. First, the device On-resistance needs to be large enough to mitigate parasitic IR drop in the array, while at the same time the conduction current in the crossbar needs to be sufficient for sensing the MVM output. It has been shown that **On-resistance** in the range of 100-200 k$\Omega$ leads to minimal IR drop for moderate crossbar array sizes (such as 64 x 64 or 128 x 128) [36]. When IR drop is inhibited, the impact of device *I-V* non-linearity is also reduced since the voltage applied on each device is close to the designed read voltage. In practice, the optimal point of ON-resistance setting may be shifted depending on the array size, as well as the sparsity in the workload [48]. Furthermore, devices with **high ON/OFF ratio and small device-to-device and cycle-to-cycle variations** are desirable for providing enough sensing margin and low noises. It has been demonstrated that ON/OFF ratio of at least 5-10 [49] is capable of implementing crossbar arrays for CIM, while higher ON/OFF ratio $10^8$-$10^9$ have been demonstrated at the device level (see Summary Table in Section 6).

The **number of bits per cell** in the memory element (in the case of digital storage of weights) dictates design choices for various factors including array density, crossbar computing error, and system-level hardware efficiency. While the demonstration of large number of bits per cell at the device level can be beneficial for providing high density on-chip storage of weights, the implications of multi-bit crossbar arrays on the hardware performance and viability needs to be examined. As we discussed before, the hardware costs of crossbar-based MVM operations are dominated by the ADC circuitry. Encoding weight matrices with many bits per cell will require increased ADC bit precision to process the crossbar output, leading to significant increase in the peripheral overhead. Based on the recent survey of ADC circuitry [50], ADC energy and area consumption will increase exponentially as bit precision increases. Therefore, performance simulations of crossbar CIM architecture have found that the crossbar arrays are optimized at about 1-2 bits (2-4 states) per memory cell in terms of hardware efficiency. Exceeding 2 bits/cell incurs significant ADC energy/area overhead that typically overshadows any benefit of memory array density. Furthermore, the programming of multi-bit arrays needs significantly longer and more complicated pulsing schemes and iterative write-and-verify steps, which may lead to additional increase in the hardware energy and latency.

## 2.6 Energy considerations

We will focus on the cross-bar array performance that will give an upper bound for the performance of the system as a whole and its implication on the material properties and operating conditions. For the best possible case of highest density, we will assume that no select devices such as, diodes or transistors are needed; this will give a specification for the fundamental material requirements. We will consider the effect of select devices later. The key operational parameters are: array size, read and write speed, operating voltage, and resistance of the switching element. We will address these for individual materials in Section 3. For a simple estimate of chip performance and chip area bounds we will have to combine hardware properties and application specific parameters.

The array speed in operations per second (Op/s) is bounded by the number of total weights $w_{tot}$ and the effective read speed which is controlled by the integration time, $t_{int}$, required to sum up the charge for each column. For a simple MVM the number of weights per layer (cross-bar) are directly related to the number of MACs that are executed in that layer. For CNN the situation is more complicated since the weight matrix is repeatedly used on the input data leading to a weight share factor ($\geq 1$) for each layer $N_i^{share}$, which has its maximum $N_{max}^{share}$ usually in the first layer [51]. As a result, the processing time per layer will be $t_{int} N_i^{share} < t_{int} N_{max}^{share}$. On the other hand, we have $MAC_{total} = \sum N_i^{share} weights_i < N_{max}^{share} weights_{total}$ and finally we obtain for the speed in operations per second

$$\frac{Op}{s} \leq \frac{2 MAC_{total}}{t_{int} * N_{max}^{share}} \leq \frac{2 weights_{total}}{t_{int}} \quad (1)$$

There are two bounds provided because the second bound is often easier to estimate with available information. Following [52] the operating voltage (V) is limited by the weight range to read noise ratio (note that we do not make any assumption on the bit accuracy as discussed above, we only assume that we can separate the signal from the noise).

$$\frac{weight\ range}{read\ noise} \sim 10 < \sqrt{\frac{V^2 t_{int}}{N R_{dev} k_B T}} \quad (2)$$

and the resistance of the switching element $R_{dev}$ has to be high enough so that the voltage-drop in the wiring levels with its resistance $r_{wire}$ (wire resistance across the unit cell) can be neglected.

$$0.1 > \frac{N^2 r_{wire}}{R_{dev}} \quad (3)$$

In addition, we need to add a lower bound for the required area that is determined by cell size of the cross-point unit cell, the tile size $Tile_{size}$ and the number of tiles $Tile_{number}$ used. Since the physical conductivities are positive, we require two conductivities per unit cell to denote signed weights. The weight is encoded as the difference of these two elements. This gives a lower bound for the cell area as four times the square of metal pitch p (~ 2F) in a planar cell configuration.

$$A > Tile_{size} Tile_{num} = 4p^2 \frac{Array_{size}}{Array_{efficiency}} Tile_{num} > 4p^2 Array_{size} Tile_{num} \quad (4)$$

In Equation (4) we have introduced the array efficiency which measures how much of the tile is covered by the actual array. This number, Array$_{efficiency}$ < 1, depends in a critical way on the unit cell design and the technology integration scheme. For the actual chip-size one has to consider the support logic and I/O circuits shown in Figure 3a as well. For the lower bound the tile number and array dimensions suffice. Using the constraints from Eq. (1) to Eq. (3) we show in Figure 6 how the operating parameters depend on the performance in operations per second (Op/s) for a system architecture consisting of 64 to 128 tiles. Since energy efficiency will scale with the square of the supply voltage, we see a critical point at about 5 POp/s at which the operating voltage has a steep increase. At the same point we observe a base resistance of about 5MΩ corresponding to a tile size of 2048x2048 cells. The number of supported weights has an upper bound of about 250M on chip. All these estimates assume an integration time at 100ns and a supply voltage margin in excess of $3k_BT$. Figure 6 illustrates clearly the interdependence of the operating parameters.

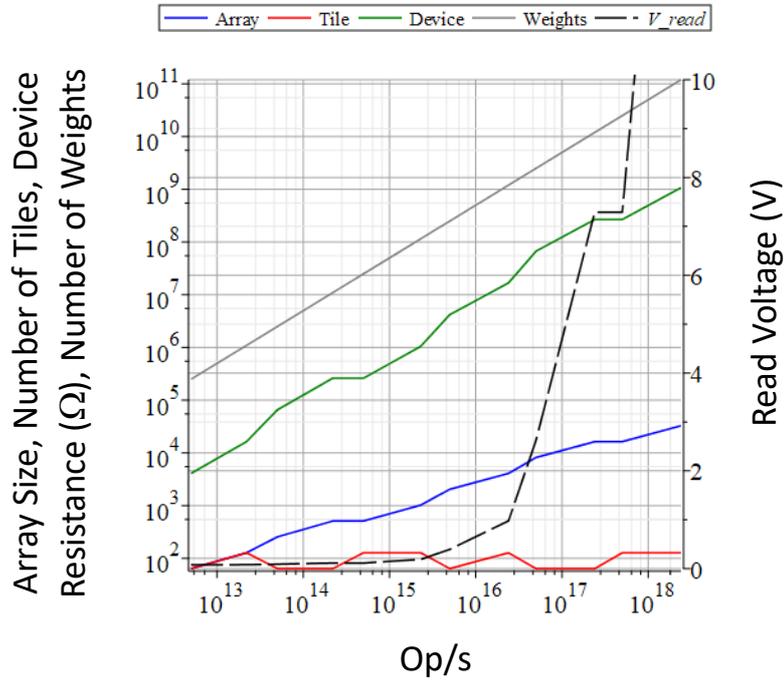

**Figure 6:** Operating parameters vs speed in operations per second. Tile number (red) is kept constant at 64 ..128 tiles. Array capacity (blue), Cross-point device resistance (green). Number of supported weights (grey). Read voltage (black dash - right vertical axis**).** Results are calculated using Equations (1) - (3) with an integration time $t_{int}$=100ns and $r_{wire}$=100mΩ.

Often demonstrations are done at smaller array sizes which is a good first step [25] [53] however, attention has to be given to the scalability of the technology to useful array sizes to accommodate the large number of weights for a relevant model. Figure 7 shows a lower bound for chip size (using Equation 4) for different array sizes (512 x 512, 1024 x 1024, 2048 x 2048 cells), shared columns per ADC (2,8,64), and metal pitch level of integration (200nm & 400nm) for a fixed number of tiles 64 and 128, respectively.

As discussed in Section 2 the ADC's will occupy a dominant share of the chip area compared to the array itself. This increases the chip area, cost, and compromises integration density. The trend to have larger and larger networks is a serious concern for the chip area to make this analog technology attractive. Array size (tile size) and tile number need to be well balanced for best performance (circuit overhead) and cost efficiency (chip size). Distributing a large network across multiple chips would have the disadvantage of reduced energy efficiency due to the fact that data has to be sent across chip boundaries. An attractive layout consideration for the future is 3D integration, that could gain integration density by stacking arrays and the ADC layers.

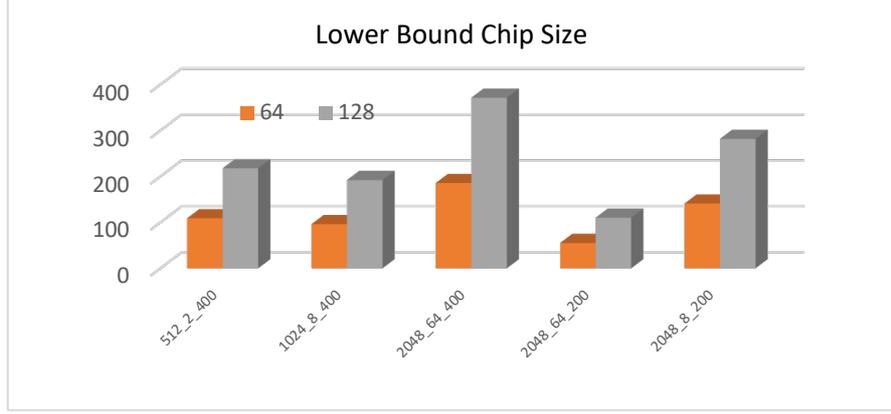

**Figure 7:** Lower bound chip size in mm² for 64 (brown) and 128 (grey) tiles. The x-axes shows the array size (512, 1024, 2048), the number of shared ADC per column (2,8 64) and the metal pitch in which the array is integrated (400nm & 200nm). Actual chip size will be larger.

To estimate the energy efficiency, we need to account for all operations analog, digital and communication. These details depend on the specific design and chip architecture and can vary significantly. However, an upper bound can be estimated by only accounting of the MACs performed in an array. For this we have

$$\text{Op/J} < \frac{2N^2}{\frac{V^2}{R_{aver}}t_{int}N^2 + E_{ADC}N} \qquad (5)$$

With V the operating voltage, $R_{aver}$ the average memory element resistance, and $E_{ADC}$ the sampling energy for the ADC and N the dimension of the weight matrix. Using Equation (5) with the bound parameters optained above we can calculate the upper bound for the energy efficiency shown in Figure 8.

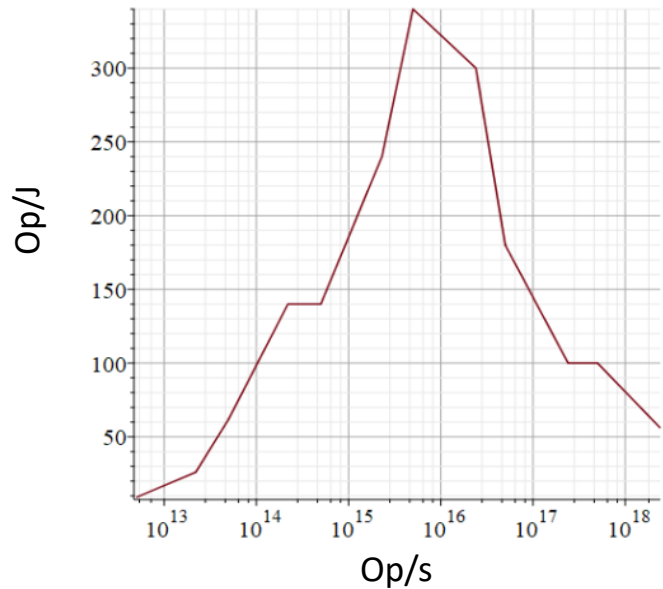

**Figure 8:** Upper bound energy efficiency in operatios per J versus operations per second. An ADC energy of 5pJ per sample is used and parameters calculated earlier (Figure 6) are used in Equation (5).

This upper bound suggests that a realistic energy efficiency will most likely fall in the range of 100 to 250 TOp/J for the system. As a comparison, note that for inference, 200TOp/J is expected to be within reach when implemented solely in CMOS technology [54]. On the other hand, for training, CMOS approaches are projected to reach only 15TOp/J in the near future [55], implying a clear energy advantage for cross-bars, though the materials requirements are more stringent (for instance write endurance) and the materials development for this purpose remains less developed. It is important to note that, due to the constant improvements in digital CMOS technology, alternative technologies to digital CMOS chase a rapidly moving target.

## 2.7 Special considerations for analog memory analog MAC

Using the memory element as an analog device in which its conductance represents the weight abandons the strictly digital representation at the possible benefit of power, area and performance. In this section we will discuss the design space and NVM requirements for this entirely analog approach which will be different than the digital memory centered analog compute approach. We will discuss a set of generic rules that can serve as guideline for the exploration of these analog components for deep learning applications [52] [51] [56] [57] and hardware solutions that may be of relevance for a manufacturable technology.

There have been an increasing number of experimental studies of all analog CIM that involve inference as well as training [58] [25] [53]. The mechanism and operation of all analog MAC has also been described in detail elsewhere [59] [52] [57]. The process of inference involves the use of NVM arrays to calculate MVM as described in Figure 3. In addition to the MVM operation in which the weights are not changed; in training a parallel operation is performed on the array that changes all weights simultaneously to perform the update, changing the conductance value of the memory elements.

We have discussed basic operational parameters as they relate to raw speed Op/s and energy efficiency Op/J. We now have to address accuracy which can be an exacting issue for all analog technologies. In the digital space this is captured by the numerical precision with which numbers are represented and computations are executed. A current safe standard would be an 8-bit representation. It is important to remember that in the digital space a number is exactly reproducible at any time in any place once the number representation is chosen as a string of 1s and 0s. The situation is very different in the analog case. A number will be represented by a conduction value that is subject to spatial and temporal variation caused by noise as well as systematic drift in the material [60]. It is preferable to talk about programming steps, between a lower and an upper conductance bound, $G_{min}$ and $G_{max}$ respectively, The conductance steps $\Delta g$ will be induced by either a voltage [58] or a current trigger [61] [62] that will change the conductance of the material.

Since for simple MVM the all-analog operation follows exactly the same process as discussed in Figure 3, we will only discuss the parallel update operation next. Recall, that the update changes the weight value incrementally. This can be achieved by injecting pulse trains (voltage or currents) of specified duration, magnitude and number of pulses from the periphery simultaneously into the vertical and horizontal metal lines of the array. At each cross-point these pulses will either coincide (add-up) or not. By choosing the appropriate magnitude of the pulse at occasions where coincidence happens the conductance will change, otherwise the conductance will stay the same. The proper design of the pulse trains will therefore allow the update to be performed in parallel. [58] [59] The polarity of the magnitude will determine if the conduction increases (potentiation) or decreases (depression). Critical for the convergence of the training algorithm is the incremental switching behavior of the memory element. In Figure 9 we show a summary of possible switching behaviors.

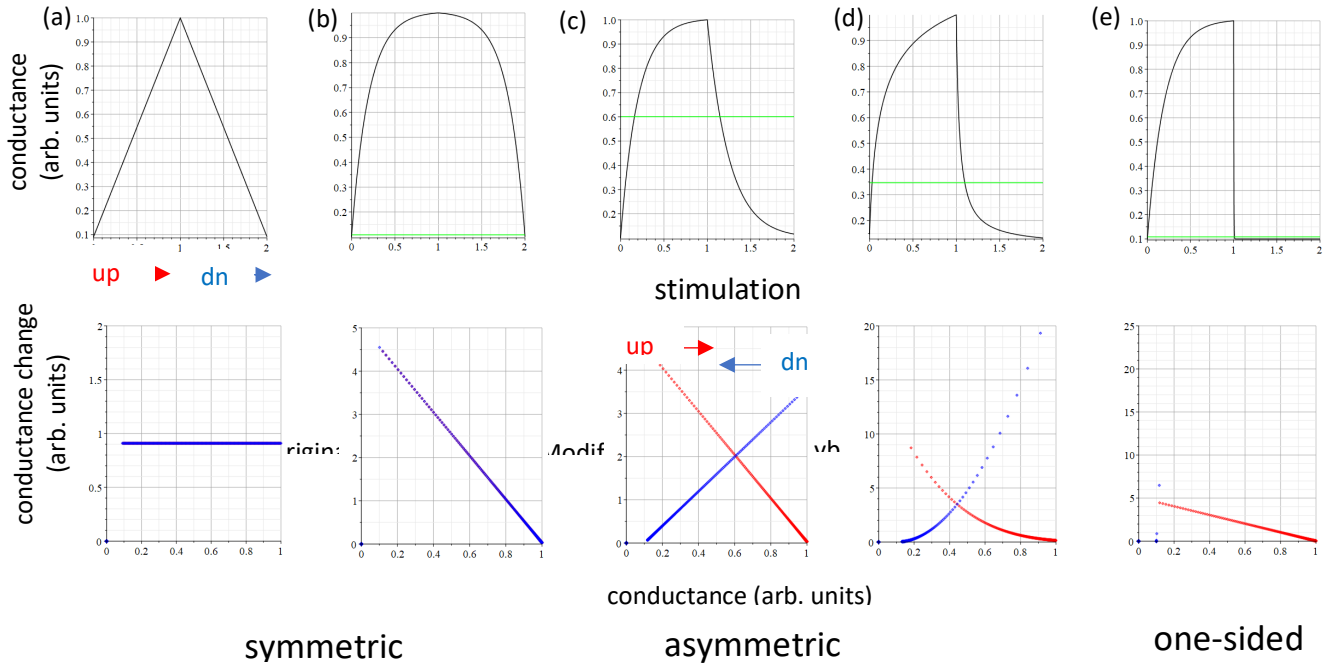

**Figure 9:** Schematics showing possible switching behavior for conductance (G) as a function of a series of potentiation (up) and depresssion(down) pulses, stimulus pulses of opposite signs, which could be either current or voltage pulses. Top row (a) thru (e) shows normalized conductance vs. pulse number. Pulses between 0 to 1 are for potentiation, and between between 1 to 2 are for depression. Bottom row (a) through (e) shows $|\Delta G|$ vs G. Note, up branch (red) runs from left to right (0..1) and down brach (blue) runs from right to left (1..0). Up and down brach for symmetric devices (a) & (b) lie on top of each other. Asymmetric devices (c) & (d) show a crossing point of up and down branch. At the crossing, ($G_{sym}$), both branches have the same $|\Delta G|$ value, they are symmetric at this point. Therefore this point is called symmetry point. The one-sided switch (e) is neither symmetric nor does it have a symmetry point.

The top row of Figure 9 shows the conductance (G) versus a sequence of a continuous "up" (potentiation) followed by a sequence of successive "down" (depression) pulses. The lower row shows the conductance change per pulse $|\Delta G|$ versus conductance G. Figures 9a and 9b show ideal cases where the conductivity profile is symmetrical for potentiation and depression. The absolute value of the derivatives as shown in the bottom row for these devices lie on top of each other, which means that for every conductance value potentiation and depression will result in equal but opposite changes. Figures 9c and 9d show devices that are asymmetric in their conductance profile for potentiation and depression and their derivatives show an intersection at a certain conductance value. At this conductance value (called symmetry point $G_{sym}$) changes in potentiation and depression are equal in magnitude but with opposite sign. Figure 9e is a case that is neither symmetrical, nor does a symmetry point exists. Note that in a device where the conductance profile is asymmetric, but a crossover point exists, a random sequence of equally balanced potentiation and depression pulses will drive the device operating around the symmetry point [63] regardless of its initial conductance. A symmetric device will under these conditions fluctuate around its initial conductance value.

The conventional back-propagation algorithm seeks to minimize the error with respect to the classification accuracy: close to convergence the weight update will be a random variation around the target. For a symmetric switching device that means the conductance will randomly vary around its algorithmic target value because up and down changes are the same at any conductance value for a symmetric switching device. Illustrated in Figure 10, the asymmetric device will move to its symmetry point and there will be a competing influence to the algorithmic target. The result is a sub-optimal solution resulting in larger classification errors.

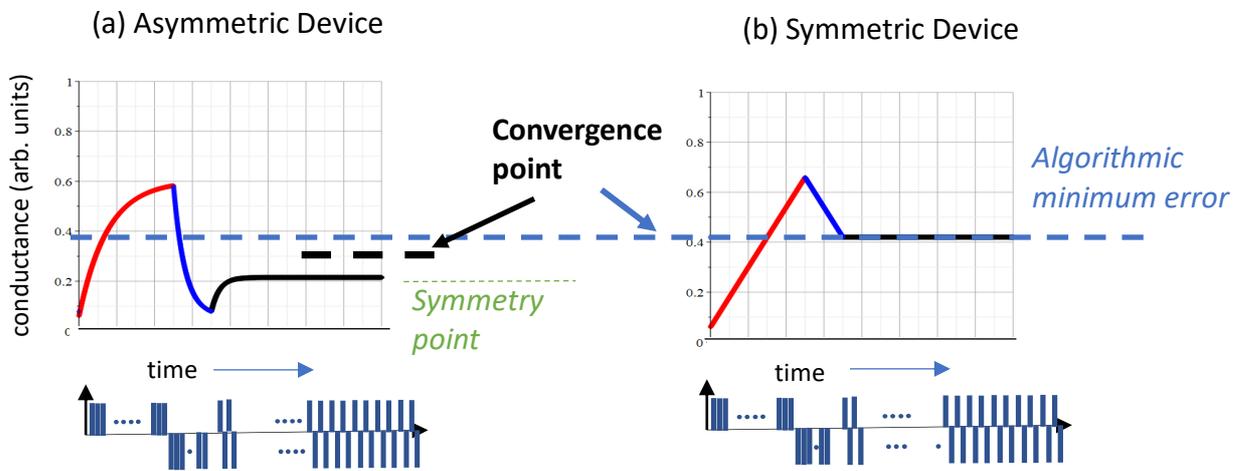

**Figure 10:** Impact of device behaviour on algorithmic convergence. At algorithmic convergence the stimulus fluctuates randomly around the stationary convergence point. (a) At the staionary point the asymmetric device has the tendnecy to move to its symmetry point therefore competing with the algorithmic target. This results in a stationary value that is off the algorthmic target and will result in a poorer classification accuracy. (b) Since up and down changes are the same for a symmetric device at any conductance value the device has no impact on the algorthmic target.

The symmetric linear switching corresponds directly to the digital case because of a constant response to stimulus. Both linear and non-linear symmetric switching elements are suited to be used in context with conventional back propagation [52]. The non-symmetric case requires a modification of the conventional back propagation algorithm [57] [64]. The modified algorithm takes advantage of the symmetry point of the device in minimizing the classification error. **This is an important (though less appreciated) point to note, since enabling the accommodation of non-ideal response in the material greatly opens up the candidate possibilities**. Consideration of performance in such algorithmic accommodation of non-idealities needs further examination. Simulations show that the modified algorithm relaxes the required number of programming steps by one order of magnitude and opens up the noise tolerance window considerably [57]. The situation changes for one-sided switching device shown in Figure 9e, which is neither symmetric nor does it possess a symmetry point. Therefore, its use for training is of limited value.

A note is in place regarding the required number of programming steps to achieve the required accuracy in the application. For the simplest case of a linear symmetric switching device the situation is clear. The maximum number of programming steps would be given by, $N_{Steps} \sim \frac{2(G_{max}-G_{min})}{\Delta G}$ where ΔG is the minimum meaningful average conductance step (a property of the device, induced by either a voltage [58] or a current trigger [61] [62] ) whose value is obtained by simulation. [47] Simulation is also required to find the effective number of programming steps for an asymmetric device. It turns out that this number is related to the conductance change around the symmetry point, $N_{Steps} \sim 2(G_{max} - G_{min})/\Delta G_{sym}$. The net of these simulation results is that for the symmetric device used in conventional back propagation the required number of programming steps is greater than 1000 at a variance of 2% for ΔG [52] [51]. For the asymmetric device used with the modified back propagation the number of programming steps is lower than 100 at a variance of 20% [57]. Although these estimates were produced including material noise and circuit constraints, the final judgement falls to hardware demonstrations on relevant applications.

We close this section with some remarks on select devices. Fundamentally the cross-point architecture for deep learning does not require select devices for all-analog CIM. There is no need to address individual cells as required in a memory. All read and write operations can be performed by peripheral control circuits in parallel for all cells simultaneously in the array. Even the weight transfer for an inference engine can be considered as a row-by-row and array-by-array (layer-by-layer) modified update mode. This, however, requires that the needed circuitry is in place. For sole inference engines that might be a burden on circuit overhead since no continuous update is required. Here the select devices provide the opportunity to run the cross-bar array as a memory with individual access to the cells to directly write the weight value. It has been shown, however, that individual direct write introduces errors that cannot easily be compensated [65]. It is preferable to transfer the weights with a modified row-by-row scheme that also includes the peripheral sense circuitry. For this solution select devices are helpful but not necessary.

## 2.8 Neurons for CIM

So far, we focused our discussions on the matrix-vector operations associated with synaptic weights. Running holistic neural network models also needs to implement neuron activations. Typical neuron activations such as sigmoid and rectified linear unit (ReLu) functions can be implemented separately in CMOS-based logic and arithmetic units. However, in order to further reduce the data movements between synapses and neurons, we can also connect neuronal functional blocks directly to the output of CIM macros. Beyond-CMOS technologies, using memristive [66], [67], spintronic devices [68], [69], metal-to-insulator transition based devices [70], and SOT-MRAM devices have been explored for emulating analog neurons. However, their development remains at a very early stage, with issues related to endurance, reliability and performance windows. It is likely that neuronal activations will continue to be digital CMOS-based in the near future, and this will not be a limiter for the introduction of analog CIM engines. That introduction today is limited by the NVM materials and the cross-bar device, which is where the focus of this review is. Once the NVM materials are sufficiently matured it is clear that activations will merge as the next challenge to improve energy efficieny for analog CIM.

# 3   Materials for Compute-in-Memory

Our discussion below will be for materials and devices for compute in-memory options (b) and (c). As already discussed, the crossbar array can contain two types of devices: the **memory element** (whether digital or analog) which sits at the junction between each crossbar interconnect, and the s**elector switch**. In some cases, these two can be integrated together. Going beyond this, and seeking more efficient non-CMOS emulations for the array periphery by the replacement of Si CMOS components, one may additionally consider **analog neurons**. For deep learning that would mean finding analog solutions for the various activation functions. The more biologically inspired spiking networks would rely on an "integrate-leak-and-fire" characteristics which would replace more complicated Si circuitry that currently mimics this function. These networks also use more sophisticated cross-point (synapse) elements with certain temporal characteristics with respect to the received stimulus. [71]. At present time neural networks are dominated by back propagation based deep learning. Genuine spiking networks have not yet entered the commercial space mainly due to a lack of accuracy.

The devices and materials treatment below will consider analog and digital memory elements for cross-bar architectures for deep learning hardware. The focus will be in relation to the needs of the embedded hardware space (described in the previous sections) within which the memory has to comply with the logic process. Many of these materials are also considered for stand-alone non-volatile memory or on-chip cache memories for digital CMOS: however here we will not address those considerations.

## 3.1   The Case for Non-volatile Memory (NVM):

Various types of memory technologies have been explored for CIM hardware. SRAM-based CIM can achieve maximized performance and speed leveraging mature CMOS technology; The SRAM cell is fast but can only store a single bit, therefore multiple (n) cells are required to store the weight at a given n-bit precision. As noted earlier the SRAM has a large cell size (160-200$F^2$). CIM hardware may require 8-10T SRAMs for increase stability (read/write disturb) and faster operation [72] [73] [74] [75] [31]. Therefore, SRAM is significantly limited in its ability to accommodate large-scale weight matrices. In addition large SRAM arrays will also contribute to the standby current of the chip because of inherent device leakage.

Consequently, there is a demand for dense, low power, on-chip memory, and various candidates have been investigated extensively over the recent years. While they will be discussed in more detail subsequently in this section, in summary the most competitive candidates at present among emerging memories are resistive random-access memory (RRAM) [76] [77] [78] [79] [80] [81] [82] [83] [84] [37], phase-change memory (PCM) [85] [34] [86], and magnetic random-access memory (MRAM) [87] [88] [89] [90] [91].

We now turn to a description of the most significant types of non-volatile memories for digital and analog cross-bars, and their applicability and challenges. We consider memory devices and materials that rely upon resistance changes. This is by far the most common type explored because they are the most compatible and adaptable with existing computational hardware. Other approaches, such as photonic memories, have been explored recently [92] [93]; however, they bring in an additional layer of complication by requiring electrical to optical conversion and shaping elements for adaptation into a digital environment that is required for additional processing. Additionally, the subject matter, and the challenges involving all-optical memories will require a separate review article.

Resistance changes can be triggered by morphology changes, filament formation, charge diffusion, interface barrier modulation and spin filtering – to name a few. The devices can be two terminal devices, which would enable the highest densities, or three terminal devices. In addition, select devices are often required to avoid parasitic currents (sneak currents) that would create an uncontrolled voltage distribution in the array during write operations. There are five different types of memories we will consider. Since there is considerable slack in the use of terminology and we clarify them below:

1. Phase Change Memory (PCM), where resistance changes are caused by reversible amorphous to crystalline transition of the memory element.
2. Memories based upon the formation of electric field driven conducting filaments, which we will refer to as Resistive Random-Access Memories (RRAM). In such memories, the resistance of the device does not change linearly as a function of device area, because of the localized nature of the filament.
3. Electrochemical Random Access Memory (ECRAM), where resistance of the device is a linear function of device area and caused by solid state electrochemical injection of charge species into the memory device across its entire electrode footprint.
4. Ferroelectric memories, which include two-terminal Ferroelectric Tunnel Junctions (FTJs) and three terminal Ferroelectric Transistors (FEFET), where device resistance is altered by electric field switchable polarization dipoles in a thin ferroelectric layer in the device stack.
5. Magnetic Random Access Memory (MRAM), where the tunneling current between two ferromagnetic metals (separated by a dielectric) can be switched between two states by altering the direction of the magnetic domains in the metals. Also in this group are materials that allow the modulation of magnetic domain sizes in a continuous fashion.

What do we seek out of such memories? Although some of the specifics depend upon specific applications (analog or digital, multilevel or 1-bit memories etc.), they share some critical performance needs. These include:

- systematic temporal stability, i.e., minimal drift of the resistance over time
- low random spatial and temporal variability
- low switching energy - low energy consumption
- high integration density (for comparison SRAM ~ $150F^2$, DRAM ~ $6F^2$, EDRAM ~ $20F^2$)
- scalability to future advanced technology nodes, and process compatibility for on-chip integration with logic.

For analog CIM we distinguish option (b) and option (c), which we described in Section 1. The material properties for each option are dictated by the architectural choices as discussed in detail in Section 2. We have shown that the material needs may differ, depending upon whether the cross-bar is used for implementing analog MVM **only** (option (b)) with digital single bit or multi-bit memory elements; or used in an all-analog approach with a sliding conductance scale, (option (c)), to implement MVM **and** weight update in the array.

Excellent recent reviews on these five types of memories, mentioned above, have been published earlier [94] [77] [95] [96] [97] [98] [99] [100] [101] [102] [103] [104]. Hence, we will not go into the details of these individual memory technologies and approaches. Rather, we will briefly describe the mechanism (for the sake of completion) and then discuss the characteristics of these materials and their applicability towards Compute-in-Memory, which is the subject of this review.

### 3.1.1 Phase Change Memory (PCM)

PCM is based on chalcogenides that have distinctive resistance depending on whether the alloy is in its crystalline or amorphous phase. A PCM device typically follows a mushroom shape cell structure connected to an underlying electrical heater component shown in Figure 11. Controlling the heat transfer process during Joule heating/cooling enables one to switch arbitrarily between these two phases. The write operations in a PCM involves Joule heating to generate amorphization and crystallization of the chalcogenide alloy. Typically, in a write operation, an abrupt melt process in the material is executed using a short and large voltage pulse through the heater; rapid cooling leads to the melt transforming to an amorphous solid while low-amplitude pulses with long duration are applied to heat up the material to reach crystallization. Partial crystallization can be reached under intermediate heating and the partially crystallized states can maintain the states after the heating is removed, providing a means to implement non-volatile multi-bit memory states or to cover a continuous range.

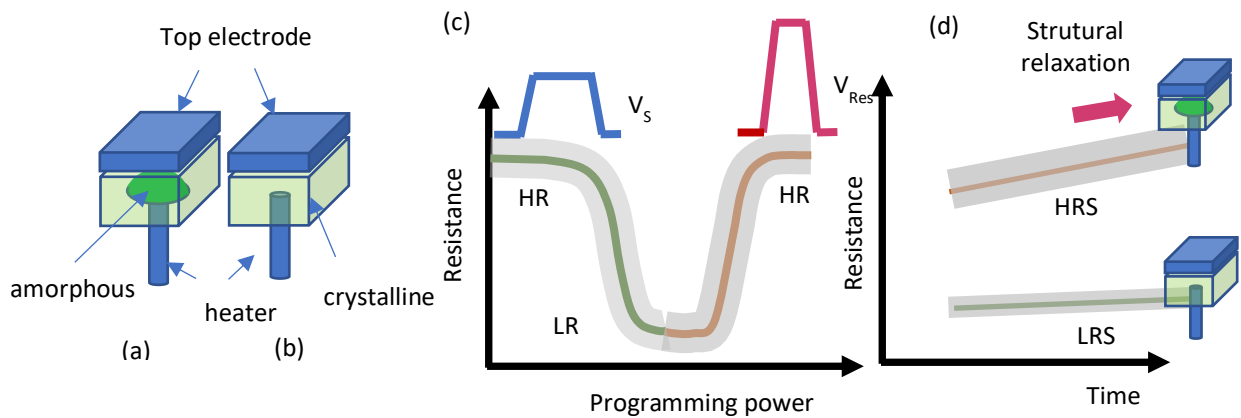

**Figure 11:** Phase change memory: (a), (b) HRS and LRS of phase change memory, respectively. The active material switches between an amorphous (high resistance) and a crystalline (low resistance) state. (c) Resistance vs. programming power for SET and RESET operations. SET operation uses a programming pulse with longer duration but consumes less power compared to RESET operation. Variations in programming (denoted by the grey region around the programming curve) can occur due to microstructural fluctuations and volume of the amorphous phase. (d) Long-term retention of PCM cell typically shows drift in the HRS towards higher resistance values. This can occur due to structural relaxations of the material in the amorphous phase

PCM has been one of the earliest explored non-volatile memories [105] and an excellent description of the mechanism, performance, and scaling characteristics of PCM devices may be found in the review by Jeyasingh et al. [106], and more recent developments in the review by Le Gallo and Sebastian [100]. The process has been shown to be scalable down to nm dimensions from a materials response perspective [107]. The large contrasts in resistance between the amorphous and the crystalline phase make phase change materials suitable for electronic data storage. Materials typically used are chalcogenide glasses, a representative compound being the ternary GeSbTe. Similar chalcogenide compounds have been used for optical storage media (compact discs) and PCM device development has benefitted from the available scientific and process infrastructure developed for optical memory in the 1990s. PCM based non-volatile memory products have existed since 2008, including use in mobile phones [108] between 2012-2014, and introduction in a cross-point memory product [109] introduced by Intel. The adaptation of PCM has not been widespread, and its use for CIM faces two intrinsically materials limited challenges. Firstly, for PCM devices the enthalpy of melting dictates the crystalline to amorphous state conversion (called the "RESET" operation) and this is a relatively high number compared to approaches that may rely upon more subtle microstructural changes for resistance change. This leads to high programming currents of ~40MA/cm$^2$, and—depending upon device size—absolute currents typically in the range of 50 mA to ~1mA. Note that typical energy costs in state-of-the-art GST-PCM cells are ~ 5 pJ and ~30 pJ for the SET and RESET operations, respectively (100 ns write latency). Creative thermal confinement strategies using interfacial layers for RESET power reduction have led to some success [110], but the need for melting the volume element essentially limits the energy efficiency of the PCM approach. Secondly, drift in the resistance, Figure 11d, of the amorphous phase has been shown to follow a power law dependence upon time following a RESET operation [111]. This limits multi-level and MVM utility because of drift in the resistance (or "synaptic") weights. Drift resilience can be built in with additional micro-architectural approaches [112], but this leads to additional complexity and energy overhead. The resistance drift arises from microstructural changes that are not clearly understood and have been related to stress relief and defect annealing over time. Frozen in, the amorphous phase can have quenched in charged point defects whose mobilities may be adequate for such drift [113] [114].

### 3.1.2 Resistive Random-Access Memory (RRAM)

RRAM is based on a simple metal-insulator-metal device structure, where the insulator can be metal oxides such as $TaO_x$, $HfO_x$, $AlO_x$ or a composite block made of multiple oxide layers. Recently, 2-dimensional (2D) materials and corresponding Van der Waals heterostructures have been incorporated in RRAMs as the switching layer [115] [116]. Halide perovskites [117] and polymers [118] have also being investigated as RRAMs in optoelectronic memory and flexible electronics applications. Metals such as Ta, Pt, Al, Cu, Ag are typically used as the electrodes sandwiching the insulator. In addition, several nanomaterials such as graphene [119] [120] and carbon nanotubes (CNT) [121] have been used as the electrode materials. The conduction mechanism in RRAMs can be described as filament formation/rupture induced by electrically driven atomic or point defect motion in the insulator. Transition from HRS to LRS occurs as conductive filament(s) form in a soft dielectric breakdown process (SET), while the reversed process (RESET) occurs when the filament is ruptured, at least partially, under an opposite electrical biasing voltage with respect to the SET operation. It has been demonstrated that RRAMs can have large dynamic range with the capability of implementing multiple bits per device. Filaments in metal-oxide RRAMs may consist of defect-rich suboxide regions, believed to have oxygen vacancies [122] [123]. This type of memory cell is referred in the literature at times as Valence Change Memory (VCM) (Figure 12 a & b). Filament formation may also occur due to fast-moving metal ions such as Ag, Cu, or Sn [124], which are supplied by one of these metals when used as an electrode (Figure 12 d & e). Such devices have been variously referred to as Electrochemical Metallization (ECM) or CBRAM though there has been loose usage of these terms in the literature. All of these devices are based upon the common phenomena of electric field driven filamentary aggregation (or disaggregation) of defects, whether they be intrinsic to the host materials, such as point defects arising from non-stoichiometry, for example oxygen vacancies [122] [123]; or extrinsic, such as injected mobile Ag or Cu ions from the electrode [124].

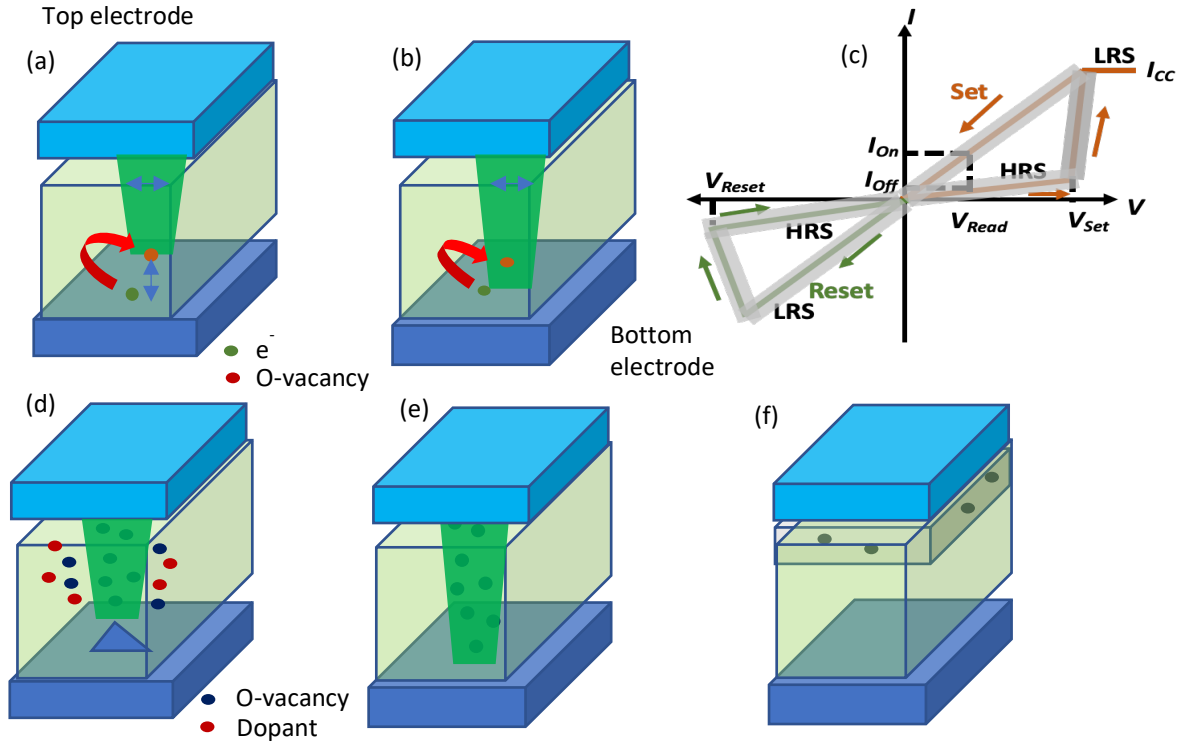

**Figure 12:** Resistive memory: (a), (b) HRS and LRS of filamentary resistive memory, respectively. The HRS state is governed by a tunneling gap, $L_g$ between the tip of the filament and the counter electrode. Variability in the device parameters can arise due to fluctuations in the filament diameter (d) and the tunneling gap. In addition, Random Telegraph Noise (RTN) and Low Frequency Noise (LFN) can originate from carrier trapping/ detrapping at trap sites such as Oxygen vacancies. (c) Typical switching *I-V* characteristics of a bipolar filamentary RRAM. A current compliance ($I_{CC}$) is used to avoid permanent dielectric breakdown. The read operation is carried out at a voltage ($V_{Read}$) much less than the SET and RESET voltages. The variations in the switching characteristics are represented by the grey regions around the switching *I-V*. (d) Strategies to reduce switching voltages: engineering the active material by addition of dopants or increasing the O-vacancy concentration within the metal-oxide. Switching / forming voltage can be also reduced by engineering asperity at the electrode. (e) Reduction of switching voltages and improvement in uniformity can be achieved by controlling filament growth through nanopores or dislocations in the switching material. (f) Interfacial resistance switching can eliminate the stochasticity involved with the filamentary mechanism. This is a bulk effect and can involve modulation of Schottky barrier at the electrode/dielectric interface.

It is important to note that the filament formation (dissolution) caused by the segregation (desegregation) of defects can be controlled by the motion of the atoms or point defects under electric drift field, as well as thermally assisted diffusion. Resistance switching is typically "bipolar", i.e., filament formation ("SET" operation) and rupture ("RESET" operation) require opposite voltage polarities (Figure 12 c). The current-voltage relationship exhibits strong non-linearity near the switching threshold. Note that the formation of the filament for the first time typically requires higher energy compared to the subsequent switching voltages. This is known as the process of "electroforming".

There has been extensive exploration of various filamentary devices in the literature for analog [78] [125] [126] [127] [33] [128] as well as digital [129] [130] [131] [132] applications. There are two material issues that are relevant to their performance. The first is the motion of the defects under electric fields which will be aligning in nature and will lead to the filament formation. Opposing this is thermal diffusion which is random and will lead to dissolution of the filament. If the electric drift component is strong and the thermal diffusion low, one has a non-volatile memory. In this case, once the filament is created, diffusion within the material is low enough at ambient temperature that the filament remains. If the components are comparable, then the filament can dissolve spontaneously upon removal of the electric field, and there is no retention of memory. One can then have a two-terminal selector switch instead of a memory. The second issue relates to the filament formation dynamics. If, at a given voltage, an increase in the size of the filament always leads to a lowering of the energy of the system (i.e. $\frac{\partial g}{\partial V} < 0$ ) where g is the free energy of the system, then this will be a runaway process till the filament shorts the dielectric layer, leading to a 1-bit memory. On the other hand, if a small increment in voltage leads to a filament growth to a point where further growth increases the energy ($\frac{\partial g}{\partial V} > 0$), then one has a steady state condition with the possibility for multiple memory states as shown in Table I.

|  | **Drift and diffusion comparable** | **Drift strong and diffusion low** |
|---|---|---|
| $\frac{\partial g}{\partial V} < 0$ | Threshold switch | Binary NV memory |
| $\frac{\partial g}{\partial V} < 0$ |  | Analog NV memory (synapse) |

**Table I:** A proposal for classification of the difference performance modes for RRAM devices based upon the free energy of filament formation

Filament driven RRAM devices are similar to PCMs in terms of their scalability, speed of operation and multi-bit storage ability. Typically, they require less RESET energy compared to PCM. However, like PCM, RRAM also suffers from variability in device to device and within device performance. Because of this, product introduction has not been widespread and are still early stage, and confined to standalone memory products. For instance, Fujitsu has recently released an 8Mb oxide-based RRAM (Fujitsu calls it RRAM) built at the 45 nm node [133] with the advantage of very low read currents.

Low voltage operation is desirable: however, it is difficult to lower voltages without affecting other parameters such as ON/OFF ratios. Reduction of forming/operating voltages can be typically achieved by engineering defects (for instance by adjusting stoichiometry of the dielectric oxide, see ref: [31] [129] in the switching layer, or doping [134] [135]). However, this leads to increased off-state leakage current, thereby reducing the dynamic range for a specific SET compliance [129]. In general, using such approaches it has generally been difficult to exceed 5 orders of magnitude in dynamic range (ON/OFF resistances) and generally high dynamic ranges are obtained at the cost of low switching voltages and vice versa. Recently, however, using nano-porous $AlO_x$ dielectric hosts (using a modified atomic layer deposition process) and with Ag as the filament forming electrode [136] high ($10^9$) ON/OFF ratios and <1V operations have been demonstrated. This was attributed to enhanced mobilities of the metal ions via surface diffusion due to the nano-porosity.

Variability in filamentary RRAM ON/OFF parameters can arise due to the stochastic fluctuations in the filament nucleation and configuration which can lead to variation in electrical response [137] [138] [139] [140]. Reducing variability via the formation of filaments repeatedly at specific locations has been explored in an epitaxial SiGe RRAM by controlled switching through a single stacking fault in the device [141]. This had led to high uniformity with 1% temporal variation in the SET voltage; however, the challenge of placing extended defects deterministically in small devices and at-scale limits its practicality.

The variability of filament-based approaches is the most significant drawback of this approach and has limited the practicability of RRAM. This has led researchers to explore changes that are areal or bulk in nature in order to reduce stochastic variations.

### 3.1.3 Electrochemical RAM (ECRAM)

In filamentary RRAMs device currents do not scale linearly with electrode area due to the localized nature of the filament. On the other hand, devices that are switched by interfacial or bulk effects are expected to scale linearly with electrode area. One such approach, ECRAM, involves a 3-terminal transistor-like device with a conductive, non-stoichiometric oxide channel that is modulated by a gate dielectric that acts as a solid-state electrolyte for ion or charged defect transport [142]., shown in Figure 13.

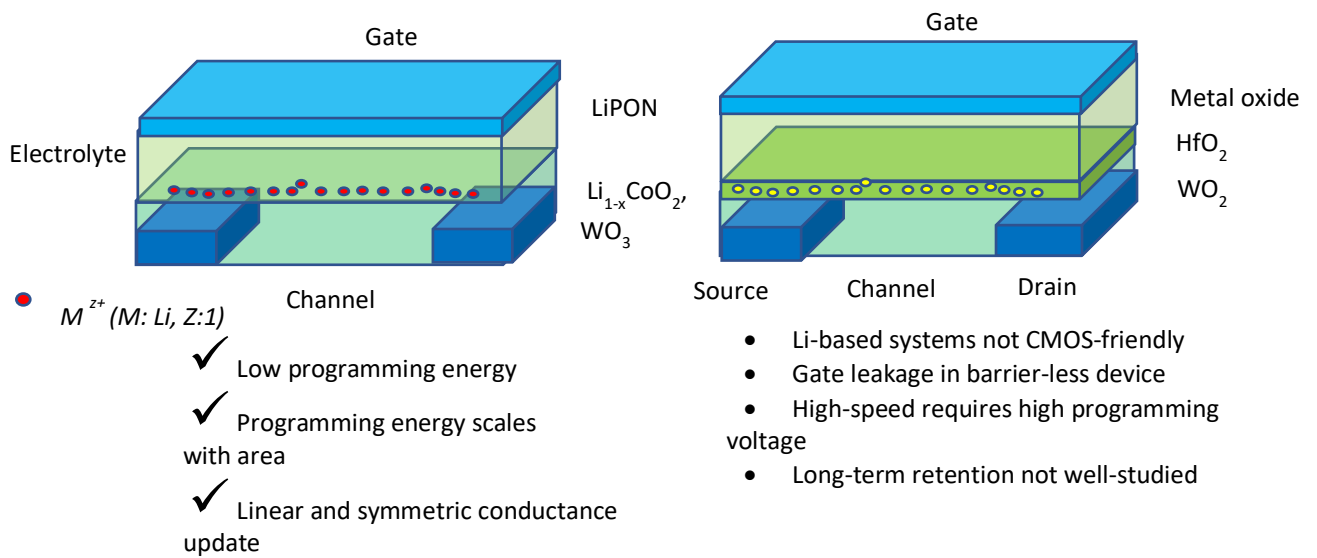

- $M^{z+}$ (M: Li, Z:1)
- ✓ Low programming energy
- ✓ Programming energy scales with area
- ✓ Linear and symmetric conductance update

- Li-based systems not CMOS-friendly
- Gate leakage in barrier-less device
- High-speed requires high programming voltage
- Long-term retention not well-studied

**Figure 13:** ECRAM: (a) Cross-sectional diagram of a typical ECRAM cell. The three-terminal configuration may consist of a channel made of $Li_{1-x}CoO_2$ or $WO_3$, an electrolyte (LiPON) and a gate/reservoir. Reversible migration of $Li^+$ ions under electric field between the electrolyte and the channel modulates the conductivity of the channel. (b) A CMOS compatible ECRAM cell comprising a $WO_3$ channel and a gate-dielectric stack of $HfO_2$ and a third metal oxide. Advantages and challenges of the ECRAM technology are shown below.

In one example (Figure 13a) the channel is $Li_xCo_{1-x}O_2$, and the gate dielectric is a solid-state electrolyte, lithium phosphorous oxynitride (LiPON). Biasing the gate leads to a change in the electrochemical potential at the LiPON- $Li_xCo_{1-x}O_2$ interface which induces Li ion exchange between the LiPON, and the $Li_xCo_{1-x}O_2$ channel and an adjustment of the channel stoichiometry on the Li sublattice. The result is a change in the $Li_xCo_{1-x}O_2$ conductivity. The IBM group [143] has pursued related devices using $WO_3$ (a conducting oxide) as the conducting channel and with both LiPON and other metal-oxide dielectric electrolytes [62] (Figure 13b). These devices are controlled by charge injection and are therefore current controlled. This is in contrast to the devices discussed before in which the filament formation is field driven and therefore voltage controlled. The variability and stability of these devices are significantly improved compared to filamentary devices. The device shows bidirectional switching and characteristics compatible with training tasks. However, as has been described by Solomon et al. [144], the devices are latency limited by slower ionic mobilities, limiting read/write speeds. Charge pileup related dynamics can lead to further problems with speed, particularly during a quick read operation after a write operation. Additionally, the current controlled nature requires a larger circuit overhead and they are therefore less suited for high density integration.

### 3.1.4 Ferroelectric tunnel junctions (FTJs) and Ferroelectric FETs (FEFETs)

In Ferroelectric tunnel junctions (FTJs) electric field driven polarization switching in a metal-FE insulator-metal stack leads to a change in the potential barrier profile across the FE dielectric layer, shown in Figure 14. This leads to a change in the magnitude of the tunneling current across the stack and indicates the memory state. The overall resistance of the device can change by orders of magnitude as a result. FTJs have a very high base resistance which can be a problem for sensing: FTJs are candidates that arrive from the high resistance side of the spectrum of candidates compared to RRAM and PCM which are candidates on the low resistance side.

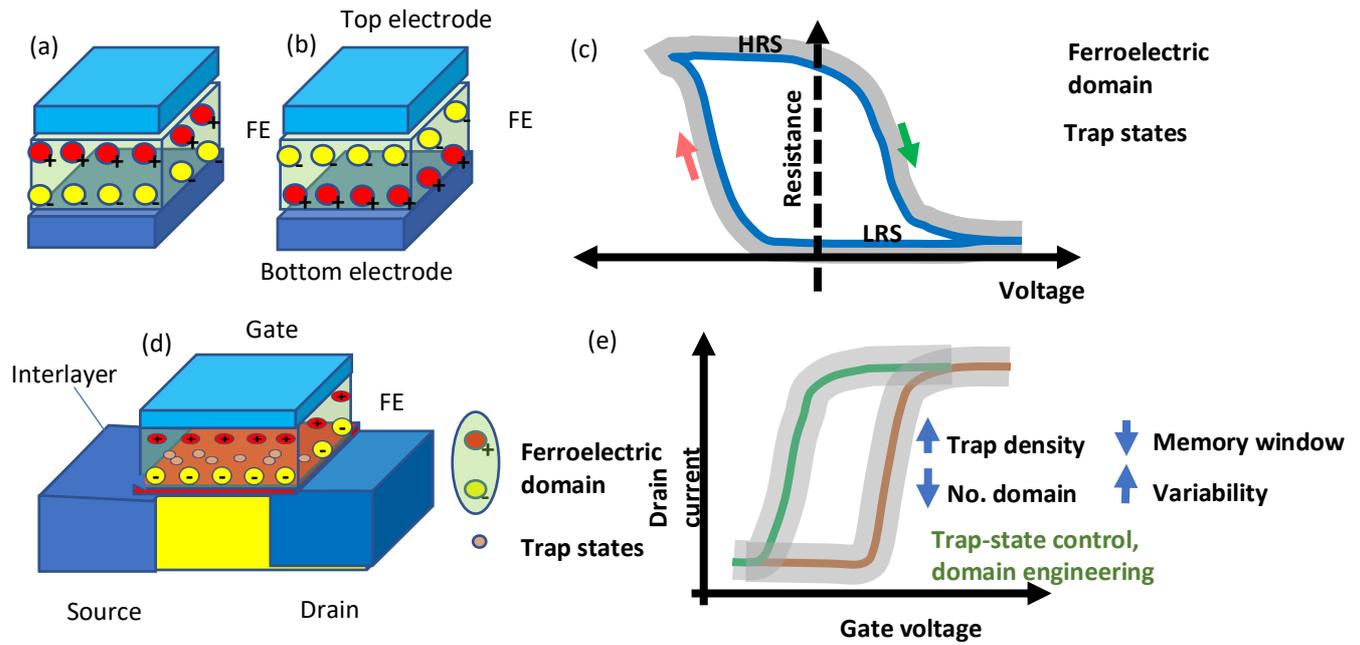

**Figure 14:** FTJ/FEFET: (a) and (b) represent two different orientations of polarization in a ferroelectric tunnel-junction. The tunneling resistance of the junction changes according the direction of polarization. (c) Evolution of resistance as a function applied voltage in an FTJ; grey region indicates variability. (d) Structure of a ferroelectric field-effect transistor (FEFET) with a ferroelectric gate dielectric (FE) and an interlayer dielectric (typically $SiO_2$). Reversal of ferroelectric polarization in the FE layer changes the threshold voltage of the transistor (e). Several factors can contribute to the variability in a FEFET device. Increase in trap density leads to a decrease in the overall memory window. Reduction in the number of ferroelectric domains have been shown to affect uniformity. Control of trap density and engineering of domain size can be some of the strategies to improve FEFET performance.

Originally, for FTJs, much of the work centered around epitaxial oxides sandwiched between a metal and a conducting oxide electrode substrate that enabled epitaxy (Figure 14 a & b). A representative example is the Co-Au / $BaTiO_3$ (BTO)/$La_{0.6}Sr_{0.33}MnO_3$ (LSMO) stack [145]. While ON/OFF ratios can be of the order $10^4$ [146], the need for an epitaxial layer and compatibility with CMOS processes limits applicability.

Discovery of ferroelectricity in doped $HfO_2$ [147] provided an alternative materials library for FTJs, obviated the need for epitaxial films and enabled CMOS compatibility. Several dopants including Si, Zr, Al, Y and La have been observed to induce ferroelectricity in $HfO_2$ films. Origin of this ferroelectricity has been identified with the emergence of the non-centrosymmetric orthorhombic phase (*Pbc*21 space group) and is believed to be related to displacement of the oxygen sublattice positions [148] [149] [150]. TiN or Pt have been used as electrodes, and while endurance and retention are good, the ON/OFF ratios for two-terminal devices have been relatively low [151], and ON currents are low, ~0.1mA/cm$^2$. This leads to poor sense margins, complicated sensing mechanisms and longer sensing times [152] [153] in considering [153] NVM applicability, and leads to more complicated circuit designs.

As a result, ferroelectric HZO films have been explored in a three terminal geometry where the FE HZO is the gate dielectric (and memory element) in a Ferroelectric Field Effect Transistor (FEFET) element (Figure 14d). Application of a gate voltage of a certain polarity leads to the ferroelectric domains getting oriented in a certain direction. Reversal of the electric field will result in ferroelectric domain growth in the opposite direction. The threshold voltage of a FeFET will change according to the direction of ferroelectric polarization in the gate dielectric. Yu et al. [153] have extensively discussed HfZO based FEFETs for synaptic applications in a recent review. The advantage of this technique is a reasonable sensing current (and therefore good read latency), and the compatibility and knowhow that already exists for $HfO_2$ based transistor technology. Additional potential advantages of this approach are fast switching (~10ns) [153] at low energies. Datta et al. [154] have reported a BEOL compatible Indium–Tungsten–Oxide (IWO) channel transistor with a HZO ferroelectric gate dielectric that showed 2 bits/cell, an endurance of $10^8$ cycles (with a memory window of ~0.5V), and a ON/OFF ratio of ~ 4. Mulaosmanovic et al. [155], using a 28 nm process flow and a Si channel transistor have shown 5 threshold voltage levels ($V_t$) levels in a single device, and controlled potentiation and depression. Global Foundries [156] have demonstrated a 22nm HZO gated Si channel FEFET, with a memory window of ~1.5V, cyclability of $10^4$ and ON/OFF ratios of a few hundred. While these are promising results, it is fair to state that the issues of cyclability, ferroelectric fatigue, device to device variability and charge trapping remain to be explored in detail. Charge trapping can lead to screening of the FE dipole, leading to drift in Vt over time and shrinkage of the memory window. Device to device variability is a second challenge, and likely related to nonuniformity in the content and distribution of the orthorhombic FE phase from device to device. This is a problem that will likely get worse with reduced dimensions. In a FEFET device, the fraction of polarization domains of opposite orientation in the gate insulator can be tuned by applying bias at the gate terminal. This will gradually shift the threshold voltage ($V_T$) of the transistor, causing either an increase or decrease in channel conductance depending upon the direction of $V_T$ shift. This can lead to "analog" memory, though charge trapping inducing drift and device to device variations limit the practicality of this specific approach.

### 3.1.5  Magnetic tunnel junctions (MRAM amd MTJ)
MRAMs utilize the relative orientation of magnetic spins to store information. The fundamental mechanism that MRAMs rely upon is the tunneling of electrons across a barrier layer sandwiched between two ferromagnetic electrodes, forming a magnetic tunnel junction (MTJ). One of the ferromagnetic layers is referred to as the fixed/hard layer which has fixed magnetization orientation, and the second layer is the switchable layer referred to as the free/soft layer as shown in Figure 15a.

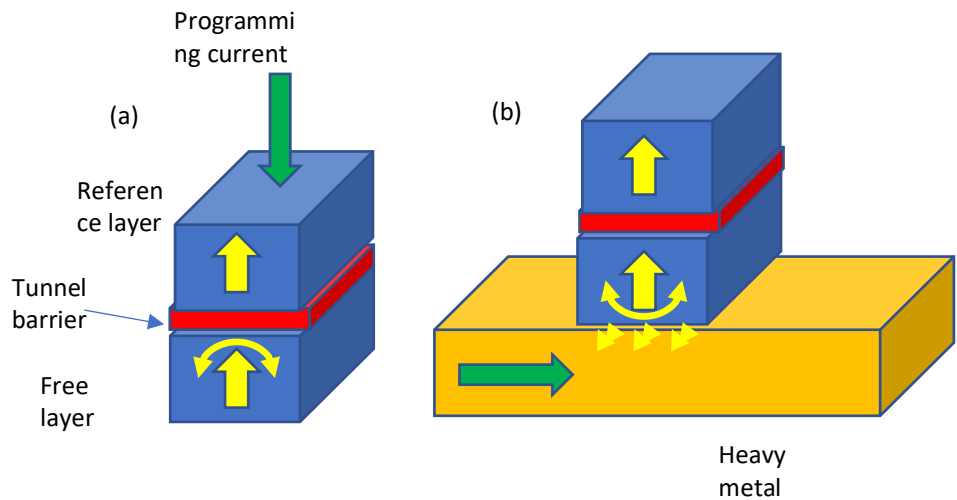

**Figure 15:** STT MRAM STT MRAM is two terminal device with write and read path are the same. Endurance concerns limitting the write current. STO MRAM is a three terminal device in which read and write path are seprated, eliminating eddurance concerns for the magnetic tunnel junction.

The high or low resistance state of the MTJ is determined by the tunneling magnetoresistance (TMR) between the two layers when they have anti-parallel or parallel magnetic orientations respectively. The TMR depends strongly on the materials used for the electrodes and barrier layer, their thickness, and interface and structure. The commonly used material currently in MRAM is CoFe-based thin films for magnetic layers and MgO as the barrier layer. TMR ratios higher than 200% have been reported in CoFeB/MgO/CoFeB MTJs [102] with reports showing a TMR of up to 600% at 300K [157] The TMR ratio increases with crystallinity of the layers and with interfacial perpendicular anisotropy due to hybridization of Fe-3d and O-2p orbitals at the interface. To have the magnetization of the fixed layer stable, typically there is an adjacent antiferromagnetic layer such as PtMn, IrMn or in recent years a synthetic antiferromagnet is used which consists of two ferromagnetic layers separated by a non-metallic spacer layer such as Ru has also been used.

Depending on the writing mechanism which switches the magnetization of the free layer, MRAMs can be classified into toggle MRAM – where an external magnetic field is used to switch the magnetization, STT/SOT MRAM – where an electric current is used to switch the magnetization. The current state-of-art MRAMs are based on the spin-transfer-torque (STT) where the electric current flows through the MTJ, and with sufficient current density, a torque is exerted on the magnetization due to transfer of spin angular momentum from the spin-polarized conduction electrons to the local magnetic moment. These are two terminal devices and typically form a 1MTJ-1T structure as shown in Figure 15a STT-MRAMs are a mature technology with several companies manufacturing and integrating these with 22 nm CMOS technology in 16 and 20 Mb memory array [158], [159], [160]. The current densities required to switch the magnetization are relatively high and impact the endurance of the device, thus putting limitations on future scaling of these devices as well as the energy cost (~ 100-1000fJ/bit) [102]. To overcome these deficiencies, an alternate mechanism relying on spin-orbit torques (SOT) has been proposed [161]. As shown in Figure 15b, the SOT-MRAM Cell is based on an MTJ placed on top of a heavy metal (HM) layer. When a transverse charge current is applied along the HM layer, an interfacial spin current will be generated flowing into the adjacent MTJ (due to the spin-orbit coupling), and the interfacial spin current can also switch the MTJ. Some of the commonly used materials for HM are Pt, Ta which give slightly lower TMR than W due to crystallographic orientation mismatch [162]. Thanks to the decoupled read/write paths, the write current of SOT-MRAM is no longer constrained by the resistance of the tunnel junction, which allows to apply large write currents in the HM layer without concerning the MTJ endurance. Enabling larger write currents will lead to faster magnetic switching, which will translate to low write latency. However, since the write path and read path are now separated, 2 transistors will be needed for controlling the read and write accesses. SOT-MRAM with 2T1R structure can have larger cell area than STT-MRAM, but it is still denser than the 6T SRAM. Therefore, SOT-MRAM can potentially reach the performance of SRAM while still providing improved on-chip density and non-volatility.

Another new developing MRAM technology is based on voltage controlled magnetic anisotropy known as VCMA-MRAM [163]. The application of electric field strongly modulates the anisotropy arising at the ferromagnet-oxide barrier layer interface thereby changing the coercive field for switching the FL magnetization. It has been demonstrated that using unipolar voltage pulses the FL magnetization can be reliably switched between two states. The significant advantage of this approach is that it uses low energy as there is minimal current required for switching, and is promising approach for scaling the devices. However, SOT-MRAM and VCMA-MRAM are still in a nascent stage of development and may require more optimization to address numerous device and material challenges before reaching maturity. Particularly, the integration of heavy metal with MTJ poses additional challenges for the device fabrication and achieving efficient deterministic SOT switching of perpendicular MTJ, or high write error rates for VCMA especially without assistance of magnetic fields remains a challenge.

Overall MRAMs offer several advantages, compared to PCM, RRAM and ECRAM, such as higher endurance, fast write speeds, and long retention times at reasonable scalability. These devices are easily integrable into a cross-bar array with several demonstrations already shown, with existing MRAM products commercially available. There have been already some demonstrations using MRAMs for neuromorphic computing. Recent work from Samsung has shown a 64 x 64 cross-bar array of STT-MRAMs integrated with 28nm CMOS technology to realize in-memory computing [164]. Work by H. Yan has also shown development of a deep-learning accelerator framework based on STT-MRAMs [165]. Newer approaches based on SOT-MRAM has also been demonstrated for accelerating graph processing applications with 5x speed up and 4x higher efficiency over DRAM based computing [166]. The integration of VCMA-MRAM with diodes can offer possibility of selector included with memory element in a cross-bar array [167].

One of the major limitations of MRAM based devices is that they are memory elements with a one-bit capacity. This limits their use for analog in-memory compute to option (b). There has been work to overcome this challenge such by using two MTJs with varying size/area to create multi-level bits. Voltage controlled domain size modulation [168], similar to ferro-electric tunnel junctions, would provide the means for multiple resistance states required for option (c). If the required granularity of about 100 steps can be achieved is currently an open question. There also still remains the problem of high write currents required for sufficient write fidelity and if the required noise margins can be reached.

## 3.2 Implementation in MVM

There have been an increasing number of NVM cross-bar implementations in the literature. The implementations have focussed on inference, which bypasses training issues related to the switching behavior of the NVM elements on one hand and the higher complexity of the peripheral circuits on the other. Embodiments have been of various complexity: from printed circuit board (PCB) assembly of peripheral circuitry with NVM array chiplets to fully integrated solutions. Performance demonstrations have typically employed the MNIST dataset of hand-written digits. These are images at an 8 bit gray scaled with 28 x 28 pixels producing an input vector of 784+1 (the 1 is added as a bias needed for the back propagation algorithm) 8 bit numbers. Often the pictures are cropped to smaller size to accommodate the hardware capacity. As a benchmark, the best software solution for this data set achieves a classification error of > 99%. Reported numbers of speed and energy efficiency of the hardware demonstrations can be difficult to interpret: a fair comparison for benchmarking needs to include all components of the system from data in to classification error out. Therefore, we will not comment on these numbers but instead focus on functionality, accuracy and identified challenges. The MVM demonstrations described below are based upon PCM, RRAM and MRAM memories.

### 3.2.1 RRAM based MVM demonstrations

In [169] a RRAM based cross-bar array with off-chip peripheral circuits was built with a Ta/HFO$_2$/Pd stack as a switching element accessible through a select transistor. Device operation was limited to the high conductance regime (100 to 700μS) to minimize non-linearities. Weight transfer was done by individually programming the cross-point devices with 64 step granularity (6-bits), and with a conductance spread of about 10μS. The demonstration vehicle was a one-layer (96 x 40) network for hand written digit classification (MNIST) with input data manipulated to fit the 96x40 array. Classification error was 89.9% which was 2.5% shy of theoretical expectation for the same configuration. Repeatability and temporal stability data were not reported. Considering the simplicity of the network the 2.5% degradation of accuracy (compared to expectation) is most likely related to inaccuracies due to the weight transfer and noise issues in the multi bit cell.

A fair assessment of RRAM NVM cross-bars for in memory compute is given in [170]. They investigated a 64 x 64 cross-bar array with Al$_2$O$_3$/TiO$_2$ RRAM cross point elements without a selector device. Weight transfer without a select device provides a challenge and the authors employed different signal tuning approaches to avoid half select disturbances during weight tuning. Weights are transferred from the learned model onto a sliding conductance scale into the RRAM elements. The paper contains a careful analysis of endurance and conductance drift behavior. Measurements show a window of 10$^5$ cycles (limited by experimental set up and not the device) and a drift of 0.7% per month with a spread estimated at 1.6% over two years. This work highlights a path to improve the yield of passive (no select device) cross-bar arrays with reduced variability and compatibility to CMOS back-end-procesing.

A more complex network implantation using RRAM MVMs is presented in [171]. This paper tackles one of the more complex issues related to in-memory computing as noted earlier: convolutional neural networks (CNN). The considered network consists of two convolution layers and one fully connected (conventional matrix) layer. The hardware implementation is based on eight 128 x 16 sub arrays mounted on a PCB board that contains all supporting peripheral logic. The array itself is a 1T1R architecture with a TiN/TaOx/HfO/Ti RRAM stack. The weights are initially determined by software training and transferred in an individual close loop write in a 2μS to 20μS conductance window with a granularity of 8 levels for positive and negative signed weights each. With 95.53% achieved accuracy the measurement falls short of the 97.99% expected theoretical accuracy for this network configuration. The imperfection of the RRAM devices is mitigated, in part, by in-situ adjusting the weights in the fully connected layer for better accuracy. Since CNNs are expected to perform better in image classification than simple fully connected networks it stands to reason that the impact of material imperfections play an important role in explaining the degraded accuracy inspite of the mitigation techniques employed.

In [172] a fully integrated solution with array and peripherals on one chip was demonstrated using a 130nm technology. This requires a NVM solution that is fully compatible with conventional CMOS processing. The unit cell was a 2T2R architecture to accommodate a signed weight value (two RRAM per unit cell) with a HfO/TaO based RRAM memory element. In the MNIST demonstration a 3bit weight quantization was used on an 784 x 100 x 10 two-layer fully connected network with a classification accuracy of 94.4%, 2% shy of the expected result. They report a low end resistance value of < 100kΩ. No information is given about the conductance operating range. No correction for device variations is reported. As discussed further below the classification accuracy will critically depend on the weight transfer and material stability which both contribute to the degraded classification accuracy.

### 3.2.2 MRAM based MVM demonstrations

The possibility of MRAM NVM has been explored in [173]. As noted earlier MRAM's limitations are that the element can only hold one bit of equivalent information, and the resistance of the device is low, limiting array size. Accuracy can be improved by bit slicing the weights as discussed in Section 2. In this early demonstration, the authors have a base array of 64 x 64 MRAM cells that was reused sequentially to meet the 784x128 and 128x10 layer dimensions for their MNIST demonstration, leading to 26 weight transfer operations per image. A one-bit accuracy for the weights was used, to avoid bit slicing. The weights were determined by software training in floating point precision initially, followed by moving to one-bit weights used in the array for inference. The reported accuracy of 93.23% falls 2% short of the expected accuracy of their network configuration which is a significant accuracy loss. Noise issues in a one-bit weight storage should be of minor importance since the memory states should be well separated. Apparently this is not the case for this demonstration which points to read out noise in the accumulation process which is another source that can degrade performance.

### 3.2.3 PCM based MVM demonstrations

More recently, two other fully integrated approaches were demonstrated [174] [175] using a PCM based cell architecture that accommodated signed weights based on multiple PCM elements per unit cell. The system was composed from tiles that contained the analog arrays and peripheral circuits for data conversion and activation. Both approaches used carefully tuned, pretrained weights that were transferred into the analog arrays. For the simple fully connected MNIST example on a two-layer network (512 x 252 x 10) they achieved classification accuracies of 97.13% and 98.3% compared to the theoretical expectations of 97.73% and 98.3%, respectively. Potential for scalability was shown in [175] in which a convolutional network with 9 layers was demonstrated using the CIFAR10 dataset. This dataset is of considerably higher complexity that the gray scale MNIST dataset and a respective classification accuracy of 85.6% was achieved compared to the expected 88.4%.

What this example highlights is that network complexity, data complexity, and hardware tuning play an essential role in the quality of the results and that a software equivalent classification error that is reached for smaller simpler networks does not neccessarily translate to more complex situations.

### 3.2.4 Overall comments regarding cross-bar demonstrations

What can we infer about the materials and device performance from these early demonstrations? A few comments in common can be made. All of the demonstrations were for inference and therefore challenges relating to training and the effects of device non-linearities and variability remain largely unexplored. For inference, the results are mostly demonstrative in nature given the early stages of this technology, and we lack adequate information about drift and long term stability. Four of the demonstrations show that the experimental results are 2-3% less accurate than theoretically anticipated. This is a significant discrepancy considering that these are relatively small cross-bars and simple data sets (grey scale no color); it points to scaling challenges to larger networks and more complex data sources. There could be various non-idealities triggering this. Low resistance of the memory element can lead to higher currents and a voltage drop from one end to the other of the metal lines in the cross-bar. This can be exacerbated for MRAM arrays due to the low resistance of MRAM (13-26k$\Omega$) which can limit its use in more complex networks. Analog noise has been ascribed as another source of the discrepancy. Discrepancies can also arise from variations in the performance of the NVM elements and extendibility to larger networks can be questionable due to noticeable impact of device variations in the loss of accuracy.

On the other hand these successful demonstrations—particularly the fully integrated solutions with the peripheral circuitry and cross-bars on the same chip—establish the compatibility of the NVM materials with CMOS process flows. The demonstrations from refs. [174] [175] using PCM devices show that it is possible to shrink the gap between the experimental cross-bar results and theoretical expectation by understanding the material issues like noise and drift and the need to devise mitigation strategies for weight transfer and circuit solutions in parallel. While non-idealities in device performance can be countered by more complex hardware designs and corrective or tuning algorithms, the overheads in energy and costs for such accommodations remain unclear. While the demonstrated results for simple networks is encouraging there is a long road ahead for this technology to compete in the market place.

## 4  Discussion

We now turn to a discussion of the relative performance characteristics of the major NVM candidates that have been used to demonstrate working cross-bar inference engines: PCM, RRAM and MRAM. We frame this discussion in reference to performance requirements from the NVM for the digital memory-analog compute, and all analog compute cases discussed in Section 2. In addition, in the summary table in Section 6 we give a snapshot of the imporatnt parameters for the NVM material classes discussed in this paper for refernce.

Note that despite the tremendous progress made over the recent decade, resistive NVMs are still not considered as a mainstream choice for memory solutions because of performance challenges that remain to be resolved. This is true for both the on-chip case (as in a cross-bar or cache application) and as standalone or external memory. In particular, and in the context of on-chip applications such as cross-bars, a large performance gap remains between most resistive NVMs and SRAM in terms of access speed, write energy and endurance. However, as has been noted earlier SRAMs are not the best fit for cross-bar memories because of their large footprint (and therefore low integration density) and also, to an extent, because of their 1-bit capacity. Table II summarizes a device comparison of SRAM and leading NVM candidates for on-chip memory applications such as cross-bars.

| Property | PCM | RRAM | MRAM | SRAM | eDRAM |
|---|---|---|---|---|---|
| Storage Density | High | High | High | Low | Medium |
| Ron/Roff | High | High | High | High | High |
| Non-volatility | Yes | Yes | Yes | No | No |
| Standby Leakage | Low | Low | Low | High | High |
| Cell Area | $16F^2$ | $16F^2$ | $30-80F^2$ | $160F^2$ | $30F^2$ |
| Write Energy | 6nJ | 2nJ | < 1nJ | < 0.1nJ | <0.2nJ |
| Write Latency | 150ns | 100ns | 10ns | < 1ns | 1-2ns |
| Endurance | $10^7$ cycles | $10^5$ cycles | $10^{15}$ cycles | > $10^{16}$ cycles | > $10^{16}$ cycles |
| Multi-Level Cell | Yes | Yes | No | No | No |

**Table II:** Comparison of NVM, SRAM, eDRAM performance for on-processor chip digital memory applications. The numbers illustrate and are broadly indicative of the qualitative differences in performance for the different materials and device approaches.

Among the NVM candidates, MRAM is the most mature, with proven endurance (>$10^{15}$ cycles), minimal drift and variability, good access time (~10-20ns), possibility of 2-4x higher on-chip density than SRAM, and at an advanced stage of development. The drawback of MRAM for crossbars is its low resistance which makes it prone to parasitic IR drop and higher power consumption; as well as the relatively low ON/OFF ratio (in the context of overall low resistance) and the binary (1 bit) storage capacity. Out-of-plane (perpendicular) spin polarized STT-MRAM [176] are currently considered the most promising candidate for embedded on-chip memory for high performance computing hardware because of the potential of higher density compared to in-plane spin polarized STT MRAM. The write latency limitation of STT-MRAM depends upon the write current density across the tunnel junction device. Low write latency and write error requires higher write currents, which impacts the endurance of the device negatively and increases power consumption. The recently developed prototype of spin-orbit-torque (SOT) MRAM [161], technology can potentially alleviate some of these issues by separating the read and write pass at the cost of cell area [48] [177].

Note that because of its 1-bit nature, MRAM will require multiple arrays to map high precision weights compared to using multi-bit RRAM or PCM. This will have an area impact and therefore the number of weights that can be implemented on a chip. Note however that there is a tradeoff—1-bit memory relieves the complexity and size of the ADC required. The low resistance of MRAM will also limit the array size due to voltage drop and power supply issues that will limit to operate the entire array in parallel. In addition, the tile area efficiency for smaller arrays decreases because for smaller arrays the ADCs will consume a relativly larger part of the tile area. Magnetic domain-controlled tunnel junctions may offer the possibility of multi-bit storage, however, their maturity has not advanced for practical use.

Compared to MRAM, RRAM and PCM offer the advantages of compact cell size, large ON/OFF ratio and multi-bit capability [78] [171] [178] [179]. However, due to the underlying conduction mechanisms in the devices, PCM and RRAM suffer reliability issues such as large device variations and retention failure due to conductance drifting and fluctuation over time [94]. In light of the variability and programmability issues with multibit devices, some recent demonstration of integrated RRAM CIM chips used binary weight storage (LRS and HRS), and partitioned high precision matrix elements to multiple memory cells [180] [181].

Compared to other candidates, PCM is energy hungry (> 10pJ) and takes longer time (150ns) to program through the heating-induced crystallization and amorphization [101]. While RRAM has a lower write energy and can be faster than PCM, it requires high programming voltages in order to activate filament formation under electrical fields. Moreover, the oxide-based devices typically have lower endurance ($10^6$-$10^9$ cycles) and suffer significant device-to-device and cycle-to-cycle variations due to the induced defects and stochasticity during the filament formation-dissociation process. In addition, both PCM and RRAM suffer conductance drift over time or under continuous write cycles, which may lead to loss of stored information.

As long as the application space is limited to inference, both PCM and RRAM are good candidates. Their high write cost is of minor concern because memory write only occurs either when the weights are transferred onto the array or at an occasional refresh to mitigate a possible systematic drift of conductance values. For most of the time these arrays will provide stationary weights that are read only during the MVM operation. Both memory elements can be used either as digital single-bit or multi-bit storage elements or in a continous range of conductances encoding the weight value. While the first two benefit from digital discretization which allows reproducibility, the latter suffers from the reproducibility of analog values. As discussed earlier, the weight transfer into an analog storage mode requires special considerations [182] [183]. Larger array sizes are a challenge for both candidates because of their high conductivity. This is a point to consider with respect to scaling to on-chip systems that can contain hundreds of millions of weights. Although both PCM and RRAM lack the endurance of MRAM, it should not be a major concern if only an occasional write occurs. Little is known how temerature effects the overall weight population and its impact on classification accuracy.

The classes of devices that are suitable for training implementation are either symmetrically switching devices with back propagation or asymmetrically switching devices that have a symmetry point with the modified back propagation algorithm. The wider class of asymmetrically switching devices comes at an algorithmic cost that requires an additional analog array [57]. PCM devices that are neither symmetric nor have a symmetry point are not natural candidates to implement training. However, PCM based training demonstrations are found in the literature [58] [25] [53].

For all-analog training, RRAM and ECRAM deserve a closer look because they allow the adjustment of their conductivity on a continous bi-directional scale which is not possible with either MRAM or PCM. Further more, RRAM and ECRAM usually have a symmetry-point that allows the implementation of the modified back-propagation algorithm. However, their low resistance range of several 10kΩ to 100kΩ prohibits the use of larger arrays to address a high weight density. The desired target resistance range around 10MΩ would provide a competitive base. Future research on these materials might benefit in approaches for increasing this base resistance. In addition, current materials are still too noisy and unstable to be useful. Temporal and spatial noise reduction would be a key problem to tackle in conjunction with increasing the resistance.

The more exploratory NVM Spin domain controlled MTJs and polarization controlled FTJs can address the resistance problem, and perhaps, enable a continous resistance range that can operate bidirectionally. Unfortunately, these materials are not mature enough to be useful. Improved reproducibility of the switching behaviour due to charge trapping (FTJ) or domain wall pinning (MTJ) will be needed to make them viable candidates for all analog CIM.

The integration density of the NVM that can be achieved is important. The more information that can be packed in a mm$^2$, the larger the network that can be stored on one chip. The minimum cell area possible would be 2 x 4F$^2$, since to support a signed weight two conductances are required. This estimate, of course, excludes select devices or other circuitry in the cell. For instance, a current or charge controlled memory element would require a current source located in each cell [184]. Among the leading NVM candidates, RRAM and PCM are the most attractive in this respect providing compact cell size (16F$^2$), and large ON/OFF ratios with multi-bit capability. The cell size by itself however, is not the only consideration. As we have discussed in Section 2 the implementation of an analog MVM engine often requires using multiple arrays to represent the weight matrix (bit slicing) which will increase the effective cell size. This increase will depend upon the number of bits stored in the memory cell—the more the number of bits/cell, then lesser the number of arrays required and the smaller the effective cell size. The effective cell size per weight is additionally determined by integration trade-offs such as placing the NVM materials in the upper layers of the metal stack on top of the select devices. As metioned earlier, depending on the complexity of the selector devices the cell-size might be dominated by the selector transistors rather the NVM element itself. Indeed, for today's implementations, the area is not constrained by the array itself but rather by the peripherial circuits, and in particular the ADCs. ADCs can occupy a significant part of the tile (Figure 3). It is particular a concern for smaller array sizes for which the ADC area occupies a relative larger portion of the tile. This is a consideration for the scalability to larger systems that involve bit slicing.

# 5 Conclusion

Considering the rapid growth of large deep learning networks, and the essential importance of the repeated matrix-vector-multiplication (MVM) operation in their execution, cross-bar MVM based compute-in-memory can be an essential hardware element of future computing architectures. They have the potential to provide significant benefits in power efficiencies and latency, particularly for training tasks, compared to CMOS. Such approaches use both digital, as well as analog synaptic non-volatile memory, and rely upon an analog multiply-and-accumulate (MAC) step that approximately scales as $2w_{total}/t_{int}$ where $w_{total}$ is the number of weights and $t_{int}$ is the integration time of the sensing circuit. Embedding an analog device element within a digital ecosystem, however, places unique challenges upon the materials and physics for non-volatile memory function, devices and circuitry. In addition, the progress of CMOS microarchitectures provides a formidable and constantly evolving moving target. Using a co-design viewpoint from microarchitectures to the materials and physics, we evaluate the status of current non-volatile memory candidates for cross-bar based CIM in terms of overall sub-system performance, scalability, energy consumption, latency and size. We do this for both analog as well as digital memory. To date, there is no candidate that fulfils all requirements, and the challenges, as well as the energy efficiency payoff benefit are more significant for training compared inference. We establish specific targets that candidate NVM materials must meet in order to be satisfactory, not from a single device performance point of view, but from the sub-system perspective that includes the cross-bar and its constituent materials as well as the sensing and analog to digital conversion. The performance of various current and emerging candidates, including those that rely upon spin angular momentum transfer (MRAM), phase change (PCM), solid-state ionics or defect (extrinsic or intrinsic) migration (RRAM, ECRAM) etc. are however, promising enough, that the requirements for the candidate materials do not seem out of reach and can bee a worthwhile challenge for materials scientists.

Acknowledgements:
This material is based upon work supported by the U.S. Department of Energy, Office of Science, for support of microelectronics research, under contract number DE-AC02-06CH11357.

# 6 Summary Table

Please note: data in table is for single device elements only

| Memory Type | PCM | RRAM | CBRAM | Interfacial RRAM | ECRAM | FEFET | FTJ | STT-MRAM |
|---|---|---|---|---|---|---|---|---|
| Mechanism and control | Morphology change heat [1-3] | Filament formation/rupture field+heat | Filament formation/rupture field+heat | charge, tunnel barrier | ionic diffusion changes channel conductance | Polarization switching of gate dielectric | Change in tunnel electroresistance | |
| Typical digital (bit capacity)/ Analog | up to 4 bits [1-3] | up to 4 bits [12-15] | up to 4 bits [23-25] | Analog [30-33] | Analog [34-37] | up to 5 bits [38] | 1 bit | 1 bit [52-55] |
| Number of terminals | 2 [1-2] | 2 [12] | 2 [18, 24] | 2 [30] | 3 [34-37] | 3 [38-45] | 2 [51-53] | 2 [52-55] |
| Min. cell size in F2 | 4 [1] | 4 [12] | 4 [18,24] | 4 | | 6 | 4 | |
| Representative switching materials | GeTe-Sb2Te3 alloys, doped Sb2Te alloys, Sb2S3, Sb2Se3 [1-2] | M-Ox (M: Hf, Al, Ti, W); [12] 2D MX2(M:Mo, W; X:S, Se); [16] 2D hBN [16] | M-Ox [18, 24], 2D-TMDC [16], 2D-nitride [16,26] | Oxide Perovskites [30] | LiPON/LiCoO2; LiPON/WO3; PEDOT:PSS/EIM:TFSI [34-37] | doped HfO2 [38-45] | BTO/LSMO, HZO [51-53] | CoFeB/MgO [52-55] |
| Representative electrodes | TiN, Al, W, CNT [1-2, 4-6] | TiN, W, Au, Pt, Ti, Al, Gr [12, 17] | Cu, Ag, Sn, CuTe, ZrTe (active); [24, 27, 28] TiN, Pt, Au (passive) [24] | Pt, Ti, Al, TiN [30 | Si (gate), Ti/Au [34-37] | TiN [38-45] | Co/Au, TiN [51-53] | |
| Separate device initialization | Initial RESET [2] | Electroforming [12] | Electroforming [24] | Yes [30] | No [34-37] | No [38-39] | No [51-53] | No |
| Nature of switching | Unipolar [3] | bipolar; unipolar [12] | bipolar; unipolar [24] | bipolar [30] | bipolar [34-37] | bipolar [38,39] | bipolar | bipolar |
| Switching non-linearity | > 1 [7] | > 1 [18] | > 1 [18] | > 1 | <1 (limited range) [34-37] | > 1 [18, 38] | > 1 [51] | |
| Typical Vthreshold/Vset | 1-2 V [2, 8] | 1-2 V [12] | < 1 V [24] | 2-4 V [30] | 0.075 - | 2-5 V [38-45] | 3-6 V [51-53] | 0.5-1.5 V [55, 62] |
| Typical $|V_{RESET}|$ | | 1-2 V [12] | < 1 V [24] | 2-4 V [30] | 0.075- | 2-5 V [38-45] | 3-6 V [51-53] | 0.5-1.5 V [55, 62] |
| Typical read voltage | < 0.5 V [2,8] | <0.5 V [20, 21] | < 0.5 V [24] | 1 V/-1V [31] | 0 V | ~ 1V [38] | 0.1 V [51] | 0.1- 0.24 V [55, 56, 62] |
| Typical write speed (standalone) | 10-100 ns [3, 8, 9] | 10 - 50 ns [12] | 10 - 50 ns [24] | $10^4 - 10^5$ ns | 5 ns - 1s [34-37] | 20 - 100 ns [38] | 20ns -50 us [51-53] | 20 - 200 ns [55-62] |
| Typical ON/OFF ratio | $10^3 - 10^5$ [1-3] | > $10^2$ [12] | $10^3 - 10^7$ [24] | $10^2 - 10^4$ | $10^2 - 10^3$ [34-37] | ~$10^2 - 10^4$ [38-44] | $10^2$ [51-53] | <10 [55-62] |
| G low (S) | 0 [10] | Passive: 0-145 uS [19] Active: 10 - 100 uS [19] | 0 [27] | 25 nS [31] | 0 [34-37] | 0 uS [38] | 14 uS [53] | |
| G high (S) | 50 uS [10] | Passive: 5- 175 uS [19] Active: 60- 900 uS [19] | 200 uS [27] | 150 nS [31] | ~300 uS [34-37] | 50 uS [38] | 35 uS [53] | |
| Typical write energy (J) | Set < 10 pJ RESET < 100 pJ [3, 9, 10] | 100 fJ - 10 pJ [18] | 1-100 pJ [27] | >100 nJ [31] | < 1fJ (<=500 nm channel) [34-37] | 1- 10 fJ [46] | ~ 2 pJ [53] | > 100 fJ [55-62] |
| Typical read energy(J) | | ~1-10 fJ [LRS]; ~0.1-1 fJ [HRS] | ~1-10 fJ [LRS]; ~0.1-1 fJ [HRS] | ~ 100 pJ [LRS]; ~10 pJ [HRS] [31] for area ~ 8x10^5 nm2 | | | ~100 fJ ~16 fJ [HRS} | |
| Typical endurance | $10^6 - 10^9$ [11] | $10^5-10^8$ [12] | $10^5-10^8$ [24] | $10^2 - 10^7$ [30] | 40 - $10^5$ [34-37] | $10^3- 10^8$ (Omega gate) [38-46] | 50 - 1000 [51-53] | Unlimited; > $10^5$ tested |
| temopral drift | Structural relaxation of amorphous phase; 1/f noise [3] | 1/f noise, RTN [20, 21] | RTN [29] | | | Depolarizing field; leakage through traps [47, 48] | | Stochastic switching due to thermal noise |
| temperature sensitivity | yes | yes | yes | yes | | yes | yes | yes |
| demonstrated dimension | < 10 nm [3, 5] | < 10 nm [22] | 10 nm [24] | 150 nm [31, 32] | 500 nm channel [34-37] | 30 nm channel (planar); 60 nm FinFET; 30 nm Omega FET [38-45] | 350 nm [51] | 15 nm [57] |
| max process temp. | < 400C [2] | < 400 C [12] | <400 C [18, 24] | <400 C [30, 31] | 400 C (MO-ECRAM) [34-37] | 500-600 C (anneal) [38] | 500-600 C (anneal) [51-53] | |

| Comments | | | Devices with operating voltages < 100 mV or operating currents < 1 nA require additional studies on long-term stability, array-level practicality | Switching speed is relatively low | Large-arrays yet to be demonstrated; additional studies on T-stability, variability and reliability necessary | MLC operation and long-term endurance can be a challenge | Large leakage and depolarizing fields; Small remnant P at ultralow thickness | |
|---|---|---|---|---|---|---|---|---|

1. G. W. Burr , JVST B, 28, 223 (2010)
2. H. S. P. Wong, Proc. IEEE 98, 2201 (2010)
3. M. Le Gallo and A. Sebastian, J. Phys. D, 53, 213002 (2020)
4. C. Ahn , Nano Lett. , 15, 6809 (2015)
5. F. Xiong, Science, 332, 568 (2011)
6. F. Xiong, Nano Lett., 13, 464 (2013)
7. A. Sebastian, J. Appl. Phys., 111101(2018)
8. M. Le Gallo, ESSDERC 2016, 373 (2016)
9. S. W. Fong, IEEE TED, 64, 4374 (2017)
10. G. W.Burr , IEEE TED, 62, 3498 (2015)
11. S. B. Kim , MRS Bulletin, 44, 710 (2019)
12. H. S. P. Wong , Proc. IEEE, 100, 1951(2012)
13. W.-C. Chien , IEDM Tech. Dig., pp. 31.5.1-31.5.4, Dec. 2011
14. A. Prakash , IEEE EDL, 36, 32 (2015)
15. S. Stathopoulos , Sci. Rep., 7, 17532 (2017)
16. T.-J. Ko , iScience, 23, p. 101676 (2020)
17. H. Fang , Adv. Electron. Mater., 3, 2017
18. S. Yu, Proc. IEEE, 106, 260 (2018)
19. Q. Xia and J. J. Yang, Nature, 18, 309 (2019)
20. S. Yu , Phys. Rev. B, 85, 045324 (2012)
21. D. Veksler , (IRPS), Apr. 2013, p. MY.10.1-MY.10.4.

22. Pi, S. , Nat. Nanotechnol. 14, 35–39 (2019)
23. K. H. Kim , Nano Lett., 12, 389 (2012)
24. D. Jana , Nanoscale Research Lett., 10, 188 (2015)
25. D. Sakellaropoulos , IEEE EDL, 41, 1013 (2020)
26. H. Zhao , Adv. Mat., 29, 1703232 (2017)
27. Y. Shi , Nat. Comm., 9, 5312 (2018)
28. S. Sonde , Nanoscale, 10, 9441 (2018)
29. R. Soni , J. Appl. Phys., 107, 024517 (2010)
30. S. Bagdzevicius , J. Electroceram., 39, 157 (2017)
31. J. jang , IEEE EDL, 36, 457-459 (2015)
32. S. Park , IEDM Tech. Dig., 2012
34. E. Fuller et. al, Adv. Mater. 2017, 29, 1604310
35. J. Tang , IEDM Tech Dig. 2018, Dec. 2018
36. P. M. Solomon , IPRS 2021, Mar. 2021, pp. 1–7.
37. S. Kim, IEDM Tech. Dig. 2019
38. M. Jerry , IEDM Tech. Dig., 2017, 6.2.1-6.2.4
39. P. Wang and S. Yu, MRS Comm., 10, 538 (2020)
40. M. Jerry , J. Appl. Phys.D, 51, 434001
41. H. Mulaomanosvi , 2017 Symp. VLSI Tech., T176-7
42. H. Mulaomanosvi , ACS Appl. Mater. Interfaces, 9, 3792-3798
43. K. Chatterjee , EDL, 2017, 38, 1379-1382

44 S. C. yan , IEEE EDL, 2021, 42, 1307-1310
45. J. H. Bae , EDL, 2020, 41, 1637-1640
46. A. I. Khan, Nat. Electron., 3, 588 (2020)
47. K. Ni , VLSI Tech. 2019
48. T. Jung , IEEE Trans. Device and Mater. Reliability, 2021, 21, 76-182
49. B. Max , ACS Appl. Electron. Mater., 2020, 2, 4023-4033
50. L. Chen , Nanoscale, 2018, 10, 15826- 15833
51. A. Chanthbouala , Nature Mat., 2012, 11, 860-864
52. W. Kim , IEDM Tech Dig. 2011
53. N. D. Rizzo , IEEE Trans. Magnetics, 2013, 49, 4441-4446
54. R. Beach , IEDM. Tech. Dig., 2008, 1-4
55. S. Chung , IEDM Tech. Dig. 2010
56. E. Kitagawa , IEDM Tech. Dig., 2012, 29.4.1-29.4.4
57. J. H. Kim , Symp. VLSI Tech., 2014
58. R. Patel , 2012 IEEE Int. SOC conference
59. H. Jarollahi , IEEE J. Emerging Sel. Top. Circuits Syst., 4, 460 (2014)
60. W. Zaho , IEEE Trans. Nanotech., 11, 907 (2012)
61. Y. Shi , IEEE EDL, 2020, 41, 1126-1129
62. S. Jung , Nature, 2022, 601, 211- 216

# 7 References

[1]  G. Boole, The Laws of Thought, Prometheus Books (2003), 1854.

[2]  J. v. Neumann, The Computer and the Brain, New Haven: Yale University Press, 1958.

[3]  A. Turing, "Intelligent Machinery," 1948.

[4]  D. Silver, A. Huang, C. J. Maddison, A. Guez, L. Sifre, G. Van Den Driessche, J. Schrittwieser, I. Antonoglou, V. Panneershelvam, M. Lanctot and S. Dieleman, "Mastering the game of Go with deep neural networks and tree search," *Nature,* vol. 529, no. 7587, pp. 484-489, 2016.

[5]  A. Krizhevsky, I. Sutskever and G. Hinton, "Imagenet classification with deep convolutional neural networks Advances in neural information processing system," in *Advances in neural information processing systems (NIPS)*, 2012.

[6]  J. Deng, W. Dong, R. Socher, L. J. Li, K. Li and F.-F. Li, " Imagenet: A large-scale hierarchical image database," in *IEEE conference on computer vision and pattern recognition*, 2009.

[7]  R. Socher, A. Perelygin, J. Wu, J. Chuang, C. Manning, A. Ng and C. Potts, "Recursive deep models for semantic compositionality over a sentiment treebank," in *Proceedings of the 2013 conference on empirical methods in natural language processing*, 2013.

[8]  A. Graves, A. R. Mohamed and G. Hinton, "Speech recognition with deep recurrent neural networks," *IEEE international conference on acoustics, speech and signal processing,* 2013.

[9]  D. Amodei, D. Hernandez, S. G, C. J, B. G and I. Sutskever, "AI and Compute," OpenAI, 2019. [Online]. Available: https://openai.com/blog/ai-and-compute/. [Accessed 2022].

[10] D. E. Rumelhart, G. E. Hinton and R. J. Williams, "Learning representations by back-propagating errors," *Nature,* vol. 323, p. pages 533–536, 1986.

[11] Y. LeCun, B. Boser, J. Denker, D. Henderson, R. Howard, W. Hubbard and L. Jackel, "Handwritten digit recognition with a back-propagation network," in *Advances in neural information processing systems*, 1989.

[12] S. Hochreiter and J. Schmidhuber, "Long short-term memory," *Neural computation,* vol. 9, no. 8, pp. 1735-1780, 1997.

[13] A. Santoro, S. Bartunov, M. Botvinick, D. Wierstra and T. Lillicrap, "Meta-learning with memory-augmented neural networks.," *In International conference on machine learning PMLR,* pp. 1842-1850, 2016.

[14] F. Scarselli, M. Gori, A. C. Tsoi, M. Hagenbuchner and G. Monfardini, " The graph neural network model," *IEEE transactions on neural networks,* vol. 20, no. 1, pp. 61-80, 2008.

[15] T. B. Brown and a. et, "Language Models are Few-Shot Learners," https://arxiv.org/pdf/2005.14165.pdf, 2020.

[16] S. Raschka, Python Machine Learning, Birmigham - Mumbai: Packt Publishing, 2015.

[17] C. C. Bell, V. Z. Han, Y. Sugawara and K. Grant, "Synaptic plasticity in a cerebellum-like structure depends on temporal order," *Nature,* vol. 387, no. 6630, pp. 278-281, 1997.

[18] A. Agrawal, A. Ankit and K. Roy, "SPARE: Spiking Neural Network Acceleration Using ROM-Embedded RAMs as In-Memory-Computation Primitives," *IEEE Transactions on Computers,* vol. 68, no. 8, pp. 1190 - 1200, 2019.

[19] J. L. Hennessy and D. A. Patterson, Computer Architecture -m A quantitative approach 6th edition, Cambridge MA: Morgan Kaufmann Publishers, 2019.

[20] J. M. Shalf and R. Leland, "Computing beyond Moore's Law," *Computer,* vol. 48, no. 12, pp. 14-23, 2015.

[21] S. Han, J. Pool, J. Tran and W. Dally, "Learning both weights and connections for efficient neural network," in *Advances in neural information processing systems*, 2015.

[22] G. Freeman and et al., "Performance-optimized gate-first 22-nm SOI technology with embedded DRAM," *Ibm Journal of Research and Development,* vol. 59, no. 1, 2015.

[23] M. Hu, H. Li, Q. Wu and G. Rose, "Hardware realization of BSB recall function using memristor crossbar arrays," in *DAC Design Automation Conference 2012*, San Francisco, 2012.

[24] M. Dazzi and et al., "Efficient Pipelined Execution of CNNs Based on In-Memory Computing and Graph Homomorphism Verification," *IEEE TRANSACTIONS ON COMPUTERS,* vol. 70, no. 6, pp. 922-935, 2021.

[25] S. Ambrogio and et al., "Equivalent-accuracy accelerated neural-network training using analogue memory," *Nature,* vol. 558, pp. 60-67, 2018.

[26] M. Hu, J. P. Strachan, Z. Li, E. M. Grafals, N. Davila, C. Graves, S. Lam, N. Ge, J. J. Yang and R. S. Williams, "Dot-product engine for neuromorphic computing: Programming 1T1M crossbar to accelerate matrix-vector multiplication," in *2016 53nd acm/edac/ieee Design Automation Conference (DAC)*, 2016.

[27] I. Chakraborty, M. Ali, A. Ankit, S. Jain, S. Roy, S. Sridharan, A. Agrawal, A. Raghunathan and K. Roy, "Resistive crossbars as approximate hardware building blocks for machine learning: Opportunities and challenges," *IEEE,* vol. 108, no. 12, pp. 2276-2310, 2020.

[28] A. Agrawal and et al., "X-SRAM: Enabling in-memory Boolean computations in CMOS static random access memories," *IEEE Transactions on Circuits and Systems ,* vol. 65, no. 12, pp. 4219-4232, 2018.

[29] M. Ali and et al., "IMAC: In-memory multi-bit multiplication and ACcumulation in 6T SRAM array," *IEEE Transactions on Circuits and Systems ,* vol. 67, no. 8, pp. 2521-2531, 2020.

[30] A. Agrawal and et al., "Xcel-RAM: Accelerating binary neural networks in high-throughput SRAM compute array," *IEEE Transactions on Circuits and Systems ,* vol. 66, no. 8, pp. 3064-3076, 2019.

[31] A. Biswas and A. P. Chandrakasan, "Conv-RAM: An energy-efficient SRAM with embedded convolution computation for low-power CNN-based machine learning applications," in *IEEE International Solid - State Circuits Conference - (ISSCC)*, 2018.

[32] A. Agrawal and et al., "CASH-RAM: Enabling in-memory computations for edge inference using charge accumulation and sharing in standard 8T-SRAM arrays," *IEEE Journal on Emerging and Selected Topics in Circuits and Systems,* vol. 10, no. 3, pp. 295-300, 2020.

[33] S. H. Jo and et al., "Nanoscale Memristor Device as Synapse in Neuromorphic Systems," *Nano Letters,* vol. 10, no. 4, pp. 1297-1301, 2010.

[34] G. W. Burr and et al., "Neuromorphic computing using non-volatile memory," *Advances in Physics X,* vol. 2, no. 1, pp. 89-124, 2017.

[35] I. Chakraborty and et al., "Pathways to efficient neuromorphic computing with non-volatile memory technologies," *Applied Physics Reviews,* vol. 7, no. 2, 2020.

[36] S. Yu, "Neuro-inspired computing with emerging nonvolatile memorys," *Proceedings of the IEEE,* vol. 106, no. 2, pp. 260-285, 2018.

[37] S. Yin and et al., "High-Throughput In-Memory Computing for Binary Deep Neural Networks With Monolithically Integrated RRAM and 90-nm CMOS," *IEEE Trans. Electron Devices,* vol. 67, no. 10, pp. 4185-4192, 2020.

[38] A. Chen, "Memory Select Devices," in *Emerging Nanoelectronic Devices, ed. An Chen, James Hytchby, Victor Zirnov and George Bourianoff*, Wiley - ISBN: 9781118447741, 2015.

[39] A. Ankit, I. E. Hajj, S. R. Chalamalasetti, G. Ndu, M. Foltin, R. S. Williams, P. Faraboschi, W. Hwu, J. P. Strachan, K. Roy and D. Milojicic, "PUMA: A programmable ultra-efficient memristor-based accelerator for machine learning inference," in *Proceedings of the Twenty-Fourth International Conference on Architectural Support for Programming Languages and Operating Systems*, 2019.

[40] A. Jaiswal and et al., "8T SRAM cell as a multibit dot-product engine for beyond von Neumann computing," *IEEE Transactions on Very Large Scale Integration (VLSI) Systems,* vol. 27, no. 11, pp. 2556-2567, 2019.

[41] A. Shafiee, A. Nag, N. Muralimanohar, R. Balasubramonian, J. P. Strachan, M. Hu, R. S. Williams and V. Srikumar, "ISAAC: A convolutional neural network accelerator with in-situ analog arithmetic in crossbars," in *ACM SIGARCH Computer Architecture News*, 2016.

[42] M. Ali and et al., "A 35.5-127.2 TOPS/W Dynamic Sparsity-Aware Reconfigurable-Precision Compute-in-Memory SRAM Macro for Machine Learning," *IEEE Solid-State Circuits Letters,* vol. 4, pp. 129-132, 2021.

[43] I. Chakraborty, A. M, K. D. E, A. A and R. K, "Geniex: A generalized approach to emulating non-ideality in memristive xbars using neural networks," in *57th ACM/IEEE Design Automation Conference (DAC)*, 2020.

[44] M. Rasch and a. et, "RAPA-ConvNets: Modified convolutional networks for accelerated training on architectures with analog arrays," *Frontiers in Neuroscience,* p. 753, 2019.

[45] S. Jain, A. Sengupta, K. Roy and A. Raghunathan, "RxNN: A framework for evaluating deep neural networks on resistive crossbars," *IEEE Transactions on Computer-Aided Design of Integrated Circuits and Systems,* vol. 40, no. 2, pp. 326-338, 2020.

[46] A. Ankit and et al., "TraNNsformer: Neural network transformation for memristive crossbar based neuromorphic system design," 2017.

[47] M. Rasch and a. et, "A flexible and fast pytorch toolkit for simulating training and inference on analog crossbar arrays," in *IEEE 3rd international conference on artificial intelligence circuits and systems (AICAS)*, 2021.

[48] T. Sharma, C. Wang, A. A and K. Roy, "Enabling robust SOT-MTJ crossbars for machine learning using sparsity-aware device-circuit co-design," in *IEEE/ACM International Symposium on Low Power Electronics and Design (ISLPED)*, 2021.

[49] F. Cai, J. M. Correll, S. H. Lee and e. al., "A fully integrated reprogrammable memristor–CMOS system for efficient multiply–accumulate operations," *Nature Electronics,* no. 2, pp. 290-299, 2019.

[50] B. Murmann, "ADC Performance Survey 1997-2021," http://web.stanford.edu/~murmann/adcsurvey.html., 2021.

[51] T. Gokmen, M. Onen and W. Haensch, "Training Deep Convolutional Neural Networks with Resistive Cross-Point Devices," *Frontiers in Neuroscience,* vol. 11, p. 538, 2017.

[52] T. Gokmen and Y. Vlasov, "Acceleration of Deep Neural Network Training with Resistive Cross-Point Devices: Design Considerations," *Frontiers in Neuroscience,* vol. 10, p. 333, 2016.

[53] S. R. Nandakumar and et al., "Mixed-Precision Deep Learning Based on Computational Memory," *Frontiers in Neuroscience,* vol. 14, p. 406, 2020.

[54] B. Zhang and et al., "A 177 TOPS/W, Capacitor-based In-Memory Computing SRAM Macro with Stepwise-Charging/Discharging DACs and Sparsity-Optimized Bitcells for 4-Bit Deep Convolutional Neural Networks," in *IEEE Custom Integrated Circuits Conference (CICC)*, 2022.

[55] S. Venkataramani and et al., "RaPiD: AI accelerator for ultra-low precision training and inference," in *2021 ACM/IEEE 48th Annual International Symposium on Computer Architecture (ISCA)*, 2021.

[56] T. Gokmen, M. Rasch and W. Haensch, "Training LSTM Networks With Resistive Cross-Point Devices," *Frontiers in Neuroscience,* vol. 12, p. 745, 2018.

[57] T. Gokmen, "Enabling Training of Neural Networks on Noisy Hardware," *Frontiers in Artificial Intelligence,* vol. 4, p. 699148, 2021.

[58] G. Burr and et al., "ExpeErimental demonstration and tolerancing of a large-scale neural network (165,000 synapses), using phase-change memory as the synaptic weight element," in *IEEE International, Electron Devices Meeting (IEDM)*, San Francisco, 2014.

[59] W. Haensch, T. Gokmen and R. Puri, "The Next Generation of Deep Learning Hardware: Analog Computing," *Proceedings of the IEEE,* vol. 107, no. 1, pp. 108-122, 2019.

[60] C. Mackin and et al., "Neuromorphic Computing with Phase Change, Device Reliability, and Variability Challenges," in *2020 IEEE International Reliability Physics Symposium (IRPS)*, 2020.

[61] E. Fuller and et al., "Li-ion synaptic transistor for low power analog computing,," *Advanced Materials,* vol. 29, no. 4, 2016.

[62] S. Kim and et al., "Metal-oxide based, CMOS-compatible ECRAM," in *IEEE International Electron Device Meeting (IEDM)*, San Francisco, 2019.

[63] H. Kim and et al., "Zero-shifting Technique for Deep Neural Network Training on Resistive Cross-point Arrays," arXiv:1907.10228, 2019.

[64] M. Onen and et al., "Neural Network Training with Asymmetric Crosspoint Elements," *Frontiers in Artificial Intelligence,* vol. 5, 2022.

[65] P. Narayanan and et al., "Fully On-Chip MAC at 14 nm Enabled by Accurate Row-Wise Programming of PCM-Based Weights and Parallel Vector-Transport in Duration-Format," *IEEE Transactions on Electron Devices,* vol. 68, no. 12, pp. 6629 - 6636, 2021.

[66] T. Tuma, A. Pantazi, G. M. Le, A. Sebastian and E. Eleftheriou, "Stochastic phase-change neurons," *Nature Nanotechnology,* vol. 693, no. 11, 2016.

[67] L. Bao, J. Kang, Y. Fang and e. al, "Artificial Shape Perception Retina Network Based on Tunable Memristive Neurons," *Scientific Reports,* vol. 8, p. 17327, 2018.

[68] A. Sengupta, Y. Shim and K. Roy, "Proposal for an all-spin artificial neural network: Emulating neural and synaptic functionalities through domain wall motion in ferromagnets.," *IEEE transactions on biomedical circuits and systems,* vol. 10, no. 6, pp. 1152-1160, 2016.

[69] S. A. Siddiqui, S. Dutta, A. Tang, L. Liu, C. Ross and B. M. A, "Magnetic domain wall based synaptic and activation function generator for neuromorphic accelerators," *Nano Letters,* vol. 20, no. 2, pp. 1033-1040, 2019.

[70] J. Lin and a. et, "Low-Voltage Artificial Neuron using Feedback Engineered Insulator-to-Metal-Transistor Devices," in *IEEE International Electron Device Meeting (IEDM)*, San Francisco, 2016.

[71] M. Davies and et al., "Loihi: A Neuromorphic Manycore Processor with On-Chip Learning," *IEEE Micro,* vol. 38, no. 1, pp. 82-99, 2018.

[72] J. Zhang, Z. Wang and N. Verma, "A machine-learning classifier implemented in a standard 6T SRAM array," in *IEEE Symposium on VLSI Circuits (VLSI-Circuits)*, 2016.

[73] Q. Dong and et al., "A 351TOPS/W and 372.4GOPS Compute-in-Memory SRAM Macro in 7nm FinFET CMOS for Machine-Learning Applications," in *IEEE International Solid- State Circuits Conference - (ISSCC)*, 2020.

[74] W. S. Khawa and et al., "A 65nm 4Kb algorithm-dependent computing-in-memory SRAM unit-macro with 2.3ns and 55.8TOPS/W fully parallel product-sum operation for binary DNN edge processors," in *IEEE International Solid - State Circuits Conference - (ISSCC), ,* 2018.

[75] S. Yin and et al., "XNOR-SRAM: In-Memory Computing SRAM Macro for Binary/Ternary Deep Neural Networks," *IEEE J. Solid-State Circuits,* vol. 55, no. 6, p. 1733–1743, 2020.

[76] D. Ielmini and H.-S. P. Wong, "In-memory computing with resistive switching devices," *Nature Electronics,* vol. 1, no. 6, pp. 333-343, 2018.

[77] Q. Xia and J. J. Yang, "Memristive crossbar arrays for brain-inspired computing," *Nature Materials,* vol. 18, no. 4, pp. 309-323, 2019.

[78] M. Prezioso and et al., "Training and operation of an integrated neuromorphic network based on metal-oxide memristors," *Nature,* vol. 521, no. 7550, pp. 61-64, 2015.

[79] P. Jain and et al., "13.2 A 3.6Mb 10.1Mb/mm2 Embedded Non-Volatile ReRAM Macro in 22nm FinFET Technology with Adaptive Forming/Set/Reset Schemes Yielding Down to 0.5V with Sensing Time of 5ns at 0.7V," in *IEEE International Solid- State Circuits Conference - (ISSCC)*, 2019.

[80] C. Li and et al., "Efficient and self-adaptive in-situ learning in multilayer memristor neural networks," *Nature Communications,* vol. 9, no. 1, p. 2385, 2018.

[81] W. H. Chen and et al., "A 65nm 1Mb nonvolatile computing-in-memory ReRAM macro with sub-16ns multiply-and-accumulate for binary DNN AI edge processors," in *IEEE International Solid - State Circuits Conference - (ISSCC)*, 2018.

[82] C. X. Xue and et al., "24.1 A A 1Mb Multibit ReRAM Computing-In-Memory Macro with 14.6ns Parallel MAC Computing Time for CNN Based AI Edge Processors," in *IEEE International Solid-State Circuits Conference - (ISSCC)*, 2019.

[83] W. He and et al., "2-Bit-Per-Cell RRAM-Based In-Memory Computing for Area-/Energy-Efficient Deep Learning," *IEEE Solid-State Circuits Letters,* vol. 3, pp. 194-197, 2020.

[84] P. Yao and et al., "Fully hardware-implemented memristor convolutional neural network," *Nature,* vol. 577, no. 7792, pp. 641-646, 2020.

[85] V. Joshi and et al., "Accurate deep neural network inference using computational phase-change memory," *Nature Communications,* vol. 11, no. 1, p. 2473, 2020.

[86] Sebastian, A, M. Le Gallo and E. Eleftheriou, "Computational phase-change memory: beyond von Neumann computing," *Journal of Applied Physics,* vol. 52, no. 44, p. 443002, 2019.

[87] S. Jain, K. Roy and A. Raghunathan, "Computing in Memory With Spin-Transfer Torque Magnetic RAM," *IEEE Transactions on Very Large Scale Integrated VLSI Systems,* vol. 26, no. 3, pp. 470-483, 2018.

[88] A. Ranjan and et al., "Approximate storage for energy efficient spintronic memories," in *52nd ACM/EDAC/IEEE Design Automation Conference (DAC)*, 2015.

[89] K. Lee and S. H. Hang, "Development of Embedded STT-MRAM for Mobile System-on-Chips," *IEEE Transactions on Magnetics,* vol. 47, no. 1, pp. 131-136, 2011.

[90] Q. Guo and et al., "AC-DIMM: associative computing with STT-MRAM," in *Proceedings of the 40th Annual International Symposium on Computer Architecture. Association for Computing Machinery*, Tel-Aviv, Israel, 2013.

[91] W. Kang and et al., "In-Memory Processing Paradigm for Bitwise Logic Operations in STT–MRAM," *IEEE Transactions on Magnetics,* vol. 53, no. 11, pp. 1-4, 2017.

[92] C. Rios and et al., "Integrated all-photonic non-volatile multi-level memory," *Nature Photonics,* vol. 9, no. 11, pp. 725-732, 2015.

[93] Y. Shen and et al., "Deep learning with coherent nanophotonic circuits," *Nature Photonics,* vol. 11, no. 7, pp. 441-446, 2017.

[94] S. Yu and et al., "Binary neural network with 16 Mb RRAM macro chip for classification and online training," in *IEEE International Electron Devices Meeting (IEDM)*, 2016.

[95] H.-S. P. Wong and et al., "Metal–Oxide RRAM," *Proceedings of the IEEE,* vol. 100, no. 6, pp. 1951-1970, 2012.

[96] J. J. Yang, D. B. Strukov and D. R. Stewart, "Memristive devices for computing," *Nature Nanotechnology,* vol. 8, no. 1, pp. 13-24, 2012.

[97] S. Slesazeck and T. Mikolajick, "Nanoscale resistive switching memory devices: a review," *Nanotechnology,* vol. 30, no. 35, p. 352003, 2019.

[98] Z. Wang and et al., "Resistive switching materials for information processing," *Nature Reviews Materials,* vol. 5, no. 3, pp. 173-195, 2020.

[99] R. Waser and M. Aono, "Nanoionics-based resistive switching memories," *Nature Materials,* vol. 6, no. 11, pp. 833-840, 2007.

[100] M. Le Gallo and A. Sebastian, "An overview of phase-change memory device physics," *Journal od Applied Physics,* vol. 53, no. 21, p. 213002, 2020.

[101] G. Burr and et al., "Phase change memory technology," *Journal of Vacuum Science & Technology B,* vol. 28, no. 2, pp. 223-262, 2010.

[102] S. Bhatti and et al., "Spintronics based random access memory: a review," *Materials Today,* pp. 530-548, 2017.

[103] H. Mulaosmanovic and et al., "Ferroelectric field-effect transistors based on HfO2: a review," *Nanotechnology,* vol. 32, no. 50, p. 502002, 2021.

[104] M. H. Park and et al., "Review and perspective on ferroelecStrukovtric HfO2-based thin films for memory applications," *MRS Communications,* vol. 8, no. 3, pp. 795-808, 2018.

[105] S. R. Ovshinsky, "Reversible Electrical Switching Phenomena in Disordered Structures," *Physical Review Letters,* vol. 21, no. 20, pp. 1450-1453, 1968.

[106] R. Jeyasingh and et al., "Phase Change Memory: Scaling and applications," in *Proceedings of the IEEE 2012 Custom Integrated Circuits Conference*, 2012.

[107] H. F. Hamann and et al., "Ultra-high-density phase-change storage and memory," *Nature Materials,* vol. 5, no. 5, pp. 383-387, 2006.

[108] P. Clarke, "Micron ships phase-change memory for Nokia phones," EETimes - https://www.eetimes.com/micron-ships-phase-change-memory-for-nokia-phones/, 2012.

[109] P. Clarke, "Intel announces optane memory M15: 3D XPoint On M.2 PCIe 3.0 x4," AnandTech - https://www.anandtech.com/show/14437/intel-announces-optane-memory-m15-3d-xpoint-on-m2-pcie-30-x4, 2019.

[110] C. Ahn and a. et, "Energy-Efficient Phase-Change Memory with Graphene as a Thermal Barrier," *Nano Letters,* vol. 15, no. 10, p. 6809–6814, 2015.

[111] M. Boniardi and et al., "Statistics of Resistance Drift Due to Structural Relaxation in Phase-Change Memory Arrays," *Transaction on Electron Devices,* vol. 57, no. 10, pp. 2690-2696, 2010.

[112] I. Boybat and et al., "Neuromorphic computing with multi-memristive synapses," *Nature Communications,* vol. 9, no. 1, p. 2514, 2018.

[113] J. Li and et al., "Explore Physical Origins of Resistance Drift in Phase Change Memory and its Implication for Drift-insensitive Materials," in *IEEE International Electron Device Meeting (IEDM)*, 2011.

[114] W. Zhang and E. Ma, "Unveiling the structural origin to control resistance drift in phase-change memory materials," *Materials Today,* vol. 41, pp. 156-176, 2020.

[115] F. Hui and et al., "2D Resistive Switching Memories: Graphene and Related Materials for Resistive Random Access Memories," *Advanvces in Electronic Materials,* vol. 3, no. 8, 2017.

[116] T. J. Ko and et al., "Two-Dimensional Near-Atom-Thickness Materials for Emerging Neuromorphic Devices and Applications," *iScience,* vol. 23, no. 11, p. 101676, 2020.

[117] J. Di and et al., "Recent advances in resistive random access memory based on lead halide perovskite," *InfoMat,* vol. 3, no. 3, pp. 293-315, 2021.

[118] W. P. Lin and et al., "Polymer-Based Resistive Memory Materials and Devices," *Advanced Materials,* vol. 26, no. 4, pp. 570-606, 2014.

[119] S. Lee and et al., "Metal oxide-resistive memory using graphene-edge electrodes," *Nature Communications,* vol. 6, no. 1, p. 8407, 2015.

[120] B. Chakrabarti, T. Roy and E. M. Vogel, "Nonlinear Switching With Ultralow Reset Power in Graphene-Insulator-Graphene Forming-Free Resistive Memories," *IEEE Electron Device Letters,* vol. 35, no. 7, pp. 750-752, 2014.

[121] C. L. Tsai and et al., "Resistive Random Access Memory Enabled by Carbon Nanotube Crossbar Electrodes," *ACS Nano,* vol. 7, no. 6, pp. 5360-5366, 2010.

[122] G. Bersuker and et al., "Metal oxide RRAM switching mechanism based on conductive filament microscopic properties," in *International Electron Devices Meeting (IEDM)*, 2013.

[123] D. H. Known and et al., "Atomic structure of conducting nanofilaments in TiO2 resistive switching memory," *Nature Nanotechnology,* vol. 5, no. 2, pp. 148-153, 2010.

[124] Y. Yang and et al., "Observation of conducting filament growth in nanoscale resistive memories," *Nature Communications,* vol. 3, no. 1, p. 732, 2012.

[125] F. M. Bayat and et al., "Implementation of multilayer perceptron network with highly uniform passive memristive crossbar circuits," *Nature Comunications,* vol. 9, no. 1, p. 2331, 2018.

[126] I. T. Wang and et al., "3D Ta/TaOx/TiO2/Ti synaptic array and linearity tuning of weight update for hardware neural network applications," *Nanotechnology,* vol. 27, no. 36, p. 365204, 2016.

[127] G. C. Adam and et al., "3-D Memristor Crossbars for Analog and Neuromorphic Computing Applications," *IEEE Transactions on Electron Devices,* vol. 64, no. 1, pp. 312-318, 2017.

[128] Z. Wang and et al., "Fully memristive neural networks for pattern classification with unsupervised learning," *Nature Electronics,* vol. 1, no. 2, pp. 137-145, 2018.

[129] Y. Chen and et al., "An Ultrathin Forming-Free $\hbox{HfO}_{x}$ Resistance Memory With Excellent Electrical Performance," *IEEE Electron Device Letters,* vol. 31, no. 12, pp. 1473-1475, 2010.

[130] S. S. Sheu and et al., "A 5ns fast write multi-level non-volatile 1 K bits RRAM memory with advance write scheme," in *Symposium on VLSI Circuits*, 2009.

[131] S. R. Lee and et al., "Multi-level switching of triple-layered TaOx RRAM with excellent reliability for storage class memory," in *Symposium on VLSI Technology (VLSIT)*, 2012.

[132] Y. B. Kim and et al., "Bi-layered RRAM with unlimited endurance and extremely uniform switching," in *Symposium on VLSI Technology* , 2011.

[133] Fujitsu, "Fujitsu Launches 12Mbit ReRAM – Largest Memory Density in ReRAM Family," https://www.fujitsu.com/jp/group/fsm/en/products/reram/spi-12m-mb85as12mt.html.

[134] B. Chakrabarti, R. V. Galatage and E. M. Vogel, "Multilevel Switching in Forming-Free Resistive Memory Devices With Atomic Layer Deposited ${\rm HfTiO}_{x}$ Nanolaminate," *IEEE Electron Device Letters,* vol. 34, no. 7, pp. 867-869, 2013.

[135] Z. Fang and et al., "$\hbox{HfO}_{x}/\hbox{TiO}_{x}/\hbox{HfO}_{x}/ \hbox{TiO}_{x}$ Multilayer-Based Forming-Free RRAM Devices With Excellent Uniformity," *IEEE Electron Device Letters,* vol. 32, no. 4, pp. 566-568, 2011.

[136] B. Chakrabarti and et al., "Nanoporous Dielectric Resistive Memories Using Sequential Infiltration Synthesis," *ACS Nano,* vol. 15, no. 3, pp. 4155-5164, 2021.

[137] R. Degraeve and et al., "Causes and consequences of the stochastic aspect of filamentary RRAM," *Insulating Films in Semiconductors,* vol. 147, pp. 171-175, 2015.

[138] T. Li and et al., "The strategies of filament control for improving the resistive switching performance," *Journal of Materials in Chemistry C,* vol. 8, no. 46, pp. 16295-16317, 2020.

[139] N. Raghavan and et al., "Stochastic variability of vacancy filament configuration in ultra-thin dielectric RRAM and its impact on OFF-state reliability," in *IEEE International Electron Devices Meeting (IEDM)*, 2013.

[140] D. Veksler and et al., "Random telegraph noise (RTN) in scaled RRAM devices," in *IEEE International Reliability Physics Symposium (IRPS)*, 2013.

[141] S. Choi and et al., "SiGe epitaxial memory for neuromorphic computing with reproducible high performance based on engineered dislocations," *Nature Materials,* vol. 17, no. 4, pp. 335-340, 2018.

[142] E. J. Fuller and et al., "Li-Ion Synaptic Transistor for Low Power Analog Computing," *Advanced Materials,* vol. 29, no. 4, p. 1604310, 2017.

[143] J. Tang and et al., "ECRAM as Scalable Synaptic Cell for High-Speed, Low-Power Neuromorphic Computing,," in *IEEE International Electron Devices Meeting (IEDM), *, 2018.

[144] P. M. Solomon and et al., "Transient Investigation of Metal-oxide based, CMOS-compatible ECRAM," in *IEEE International Reliability Physics Symposium (IRPS)*, 2021.

[145] A. Chanthbouala and et al., "A ferroelectric memristor," *Nature Materials,* vol. 11, no. 10, pp. 860-864, 2012.

[146] V. Garcia and M. Bibes, "Ferroelectric tunnel junctions for information storage and processing," *Nature Communications,* vol. 5, no. 1, p. 4289, 2014.

[147] T. S. Böscke and et al., "Ferroelectricity in hafnium oxide thin films," *Applied Physics Letters,* vol. 99, no. 10, p. 102903, 2011.

[148] R. Materlik, C. Künneth and A. Kersch, "The origin of ferroelectricity in Hf1−xZrxO2: A computational investigation and a surface energy model," *Journal of Applied Physics,* vol. 117, no. 13, p. 134109, 2015.

[149] X. Sang and et al., "On the structural origins of ferroelectricity in HfO2 thin films," *Applied Physics Letters,* vol. 106, no. 16, p. 162905, 2015.

[150] J. Müller and et al., "Ferroelectric Hafnium Oxide Based Materials and Devices: Assessment of Current Status and Future Prospects," *ECS Journal Solid State Science & Technology,* vol. 4, no. 5, pp. N30-N35, 2015.

[151] B. Max and et al., "Hafnia-Based Double-Layer Ferroelectric Tunnel Junctions as Artificial Synapses for Neuromorphic Computing,," *ACS Applied Electronics Materials,* vol. 2, no. 12, pp. 4023-4033, 2020.

[152] Y. Luo and et al., "Benchmark of Ferroelectric Transistor-Based Hybrid Precision Synapse for Neural Network Accelerator," *IEEE Journal o Explorative Solid-State Computing Devices & Circuits,* vol. 5, no. 2, pp. 142-150, 2019.

[153] P. Wang and S. Yu, "Ferroelectric devices and circuits for neuro-inspired computing," *MRS Communications,* vol. 10, no. 4, pp. 538-548, 2020.

[154] K. A. Aabrar and et al., "BEOL-Compatible Superlattice FEFET Analog Synapse With Improved Linearity and Symmetry of Weight Update," *IEEE Transactions on Electron Devices,* vol. 69, no. 4, pp. 2094-2100, 2022.

[155] M. Trentzsch and et al., "A 28nm HKMG super low power embedded NVM technology based on ferroelectric FETs," in *IEEE International Electron Devices Meeting (IEDM*, 2016.

[156] S. Dünkel and et al., "A FeFET based super-low-power ultra-fast embedded NVM technology for 22nm FDSOI and beyond," in *IEEE International Electron Devices Meeting (IEDM)*, 2017.

[157] S. Ikeda and et al., "Tunnel magnetoresistance of 604% at 300K by supression of Ta diffusion in CoFeB/MgO/CoFeB psuedo-spin-valves annealed at high temperature," *Applied Physics Letters,* vol. 93, p. 082508, 2008.

[158] O. Golonzka and et al., "MRAM as Embedded Non-Volatile Memory Solution for 22FFL FinFET Technology," *IEEE Int. Electron Devices Meeting,* pp. 18.1.1-18.1.4, 2018.

[159] e. a. V. B. Naik, "Manufacturable 22nm FD-SOI Embedded MRAM Technology for Industrial-grade MCU and IOT Applications," *IEEE Int. Electron Devices Meeting,* 2019.

[160] e. a. Y. Song, "emonstration of Highly Manufacturable STT-MRAM Embedded in 28nm Logic," *IEEE Int. Electron Devices Meeting.,* 2018.

[161] K. Garello and et al., "SOT-MRAM 300MM Integration for Low Power and Ultrafast Embedded Memories," in *IEEE Symposium on VLSI Circuits*, 2018.

[162] S. J. Lin and et al., "Challenges toward Low-Power SOT-MRAM," *IEEE International Reliability Physics Symposium,* pp. 1-7, 2021.

[163] J. Lourenbam and J. Huang, Electric-Field-Controlled MRAM: Physics and Applications, Singapore: Springer, 2021.

[164] S. Jung and et al., "A crossbar array of magnetoresistive memory devices for in-memory computing," *Nature,* pp. 211-216, 2021.

[165] H. Yan and et al., "iCELIA: A Full-Stack Framework for STT-MRAM-Based Deep Learning Acceleration," *IEEE Transactions on Parallel and Distributed Systems,* pp. 408-422, 2020.

[166] S. Angizi, J. Sun, W. Zhang and D. Fan, "GraphS: A Graph Processing Accelerator Leveraging SOT-MRAM," in *2019 Design, Automation & Test in Europe Conference & Exhibition*, 2019.

[167] R. Dorrance and et al., "Diode-MTJ Crossbar Memory Cell Using Voltage-Induced Unipolar Switching for High-Density MRAM," *IEEE Electron Device Letters,* pp. 753-755, 2013.

[168] S. Lequeux and et al., "A magnetic synapse: multilevel spin-torque memristor with perpendicular anisotropy," *Scientific Reports,* p. 31510, 2016.

[169] M. Hu and et al., "Memristor-Based Analog Computation and Neural Network," *Advanced Materials - Communication,* vol. 30, no. 9, 2018.

[170] H. Kim and et al., "4K-memristor analog-grade passive," *Nature Communications,* vol. 12, p. 5198, 2021.

[171] P. Yao and et al., "Fully hardware-implemented memristor convolutional neural network," *Nature,* vol. 577, p. 641–646, 2020.

[172] Q. Liu and et al., "A Fully Integrated Analog ReRAM Based 78.4TOPS/W," in *IEEE International Conference on Solid-State Circuits (ISSCC)*, 2020.

[173] S. Jung and et al., "A crossbar array of magnetoresistive," *Nature,* vol. 601, p. 211–216, 2022.

[174] P. Narayanan and et al., "Fully on-chip MAC at 14nm enabled by accurate row-wise programming of PCM-based weights and parallel vector-transport in duration-format," in *Symposium on VLSI Technology*, 2021.

[175] R. Khaddam-Aljameh and et al., "HERMES Core–A 14nm CMOS and PCM-based In-Memory Compute Core using an array of 300ps/LSB Linearized CCO-based ADCs and local digital processing," in *Symposium on VLSI Circuits*, 2021.

[176] G. Hu and et al., "Spin-transfer torque MRAM with reliable 2 ns writing for last level cache applications," in *IEEE International Electron Devices Meeting (IEDM)*, San Francisco, 2019.

[177] J. Doevenspeck and et al., "SOT-MRAM based analog in-memory computing for DNN inference," in *IEEE Symposium on VLSI*, 2020.

[178] P. Yao and et al., "Face classification using electronic synapses," *Nature Communications,* vol. 8, no. 1, p. 15199, 2017.

[179] O. Bichler and et al., "Visual Pattern Extraction Using Energy-Efficient '2-PCM Synapse' Neuromorphic Architecture," *IEEE Transactions on Electron Devices,* vol. 59, no. 8, pp. 2206-2214, 2012.

[180] C.-X. Xue, "24.1 A 1Mb Multibit ReRAM Computing-In-Memory Macro with 14.6ns Parallel MAC Computing Time for CNN Based AI Edge Processors," in *International Solid- State Circuits Conference*, 2019.

[181] F. Cai and et al., "A fully integrated reprogrammable memristor–CMOS system for efficient multiply–accumulate operation," *Nature Electronics,* pp. 2, 290–299, 2019.

[182] C. Mackin and et al., "Optimized Weight Programming for Analogue Memory-based Deep Neural Networks," Research Square - DOI: https://doi.org/10.21203/rs.3.rs-1028668/v1, 2021.

[183] S. Nandakumar and et al., "Precision of synaptic weights programmed in phase-change memory devices for deep learning inference," in *IEEE International Electron Devices Meeting (IEDM*, San Francisco, 2020.

[184] Y. Kohda and et al., "Unassisted true analog neural network training chip," in *IEEE International Electron Devices Meeting (IEDM)*, San Francisco, 2020.

[185] S. Roy, S. Sridharan, S. Jain and A. Raghunathan, "TxSim: Modeling training of deep neural networks on resistive crossbar systems," *IEEE Transactions on Very Large Scale Integration (VLSI) Systems,* vol. 29, no. 4, pp. 730-738, 2021.